\documentclass{article}
\usepackage{arxiv}

\usepackage{cite}
\usepackage{amsmath,amssymb,amsfonts}
\usepackage{graphicx}
\usepackage{textcomp}
\usepackage{xcolor}
\usepackage[ruled,vlined]{algorithm2e}
\usepackage{tabularx}
\usepackage{longtable}
\usepackage{nicefrac}
\usepackage{hyperref}
\usepackage{subcaption}
\usepackage{microtype}
\SetKwInput{KwInput}{Input}
\SetKwInput{KwOutput}{Output}
\SetKw{ForAll}{for all}
\SetKw{Function}{Function}
\SetKw{Comment}{// }

\SetCommentSty{mycommfont}

\def\BibTeX{{\rm B\kern-.05em{\sc i\kern-.025em b}\kern-.08em
    T\kern-.1667em\lower.7ex\hbox{E}\kern-.125emX}}

\newtheorem{definition}{Definition}

\title{NI-ORCA: A Parallel Algorithm for Counting the Orbits of Non-Induced Graphlets up to \texorpdfstring{$K_4$}{K4}}

\author{
 Syed Ibtisam Tauhidi \\
  Queen's University Belfast\\
  Northern Ireland, UK BT9 5BN \\
   \And
 Arindam Karmakar \\
  Tezpur University\\
  Assam, India 784028 \\
  \And
 Thai Son Mai \\
  Queen's University Belfast\\
  Northern Ireland, UK BT9 5BN \\
  \And
 Hans Vandierendonck \\
  Queen's University Belfast\\
  Northern Ireland, UK BT9 5BN \\
}

\begin{document}
\maketitle
\begin{abstract}
Counting the orbits of graphlets in a network is a vital tool for understanding the structural roles of vertices in various graph analytics tasks. While existing algorithms efficiently compute orbits of induced graphlets, many real-world applications require non-induced orbit counts. However, no current method offers exact, scalable, and parallel support for non-induced orbit counting. This paper presents NI-ORCA, a parallel algorithm to efficiently compute the orbits of non-induced graphlets up to size four (4-clique). NI-ORCA extends the ORCA framework for non-induced orbit counting by reformulating a system of linear equations. The algorithm consists of three stages: triangle counting, 4-clique enumeration, and orbit solving. We design and implement stage-specific parallelisation strategies using thread and vertex-local memory models and data structures, minimising contention and balancing workload. We further analyse the impact of scheduling policies, chunk sizes, and affinity strategies on performance. Experimental analysis on eight real-world datasets and a series of synthetic Erdős–Rényi graphs demonstrates that a mixed mode combining stage-specific data structure, with dynamic scheduling with small chunk sizes, delivers consistent speedup and effective load balancing. Our results show that NI-ORCA significantly outperforms state-of-the-art sequential algorithms, achieving up to 30× speedups.
\end{abstract}


\section{Introduction}
Graphs, or networks, are ubiquitously used data structures in computer science, and are used to naturally represent data across a wide array of domains, including biology, astrophysics, cheminformatics, logistics, and social or computer networks~\cite{snapnets,nr}. A key task in graph analytics is understanding the structural role of individual vertices within the graph, characterising them using various descriptive parameters. To extract insights from such graph data, it is often necessary to describe the structure of a graph using various topological features. Prior studies have shown that analysing the global and local topology of graphs enables richer characterisation for tasks like classification, clustering, and similarity comparison~\cite{tauhidi2023machine,zhang2021motif,zhang2024motif,li2021analyzing}. One practical approach for this is counting the orbits of graphlets, which quantifies how frequently~vertices~participate~in~specific topological roles in the graph.

Graphlets~\textemdash~also known as \textit{motifs}~\textemdash~are small induced or non-induced subgraphs embedded within a larger graph~\cite{ahmed2015efficient}. However, \textit{motifs} often denote the graphlets that occur more frequently (e.g., based on z-score or p-value) within a graph~\cite{martorana2024motif,li2021investigating}. They capture recurring local patterns in graph topology and have been widely used in applications such as network comparison, node classification, clustering, anomaly detection, and biological function prediction. A closely related concept is that of the \textit{orbits} of graphlets. Orbits are automorphic groups of vertices in a graphlet, i.e., vertices belonging to the same orbit can be mapped to one another in an isomorphism of a graphlet to itself~\cite{pashanasangi2020efficiently}. Orbits~capture~the~specific~roles~of vertices within these graphlets.

Characterising a graph using orbit counts has several applications, such as role-based node analysis, machine learning, and network alignment and comparison~\cite{feng2020link,roux2023graphlet,ribeiro2021survey}. Additionally, orbit-based features preserve fine-grained topological signatures while remaining computationally tractable for small graphlets.  For these reasons, efficient and scalable algorithms for orbit counting have become a major area of interest. While several existing works focus on counting the orbits of induced graphlets, there is a gap in studying the algorithms for the scalable counting of orbits of non-induced graphlets. Algorithms like ORCA~\cite{hovcevar2014combinatorial}, EVOKE~\cite{pashanasangi2020efficiently}, and JESSE~\cite{melckenbeeck2016algorithm} are examples of the former. These methods use combinatorial techniques, graph decomposition, or equation systems to compute orbit counts, often avoiding full enumeration for efficiency. However, they are designed for counting the orbits of induced graphlets and are sequential.

Many applications prefer the orbit count of \textit{non-induced} graphlets instead. For example, non-induced graphlet orbits are crucial for analysing noisy or incomplete protein–protein interaction (PPI) networks, as they tolerate missing or spurious edges and still capture meaningful structural motifs. Unlike induced orbits, which require exact subgraph matches, non-induced orbits can reveal biologically relevant patterns even when some interactions are absent~\cite{luo2014combimotif}. Furthermore, recent work has shown that using non-induced orbit counts can improve performance in algorithmic tasks such as subgraph isomorphism search by significantly pruning the search space~\cite{tauhidi2024orbitsi}. Despite this, there has been a lack of exact, efficient and scalable algorithms for counting orbits of non-induced graphlets.




To address these challenges, we propose \textbf{NI-ORCA}, an exact and parallel algorithm designed to count non-induced graphlet orbits up to size four ($K_4$). NI-ORCA builds on ORCA but introduces support for non-induced graphlets and leverages modern parallel architectures. At its core, the algorithm reuses the combinatorial equation framework from ORCA, but replaces the subtractive corrections required in the induced case. Both algorithms are structured into three stages. Stage I counts triangle participation for each edge. Stage II identifies 4-cliques ($K_4$) using efficient neighbourhood intersection techniques. Stage III computes the orbit counts for each vertex using a set of internal structural variables and a system of linear equations derived from local substructures.

A major contribution of our work is the design of parallelisation strategies for each algorithm stage. While Stage I is trivially parallelisable, Stages II and III present significant challenges due to data dependencies and memory contention. We address these by introducing various thread-safe data structure designs, including per-thread and per-vertex arrays, unordered maps (UOM), and cache-friendly flat hash maps (FHM). We explore multiple configurations for these data structures to minimise contention and maximise memory locality. In particular, a hybrid \texttt{MIXED} mode~\textemdash~using thread-local FHM for Stage II and atomic arrays for Stage III~\textemdash~achieves the best performance in our evaluation. We also explore the impact of scheduling policies and chunk sizes in OpenMP-based execution. Our experiments show that dynamic scheduling with small chunk sizes provides the most consistent speedup and~best~load~balancing~across~threads.

We evaluate NI-ORCA on eight real-world networks ranging from small biological graphs like YEAST and HPRD to larger social and citation networks like YOUTUBE, EU2005, and PATENTS. Additionally, we generate synthetic Erdős–Rényi graphs with controlled density and size to study runtime characteristics in a more systematic manner. Our results demonstrate that the parallel NI-ORCA algorithm consistently outperforms state-of-the-art baselines in terms of speedup. Additionally, it demonstrates efficient scalability with increasing thread counts, on both AMD EPYC and Intel XEON multi-core, multi-socket shared-memory systems.

\vspace{0.5cm}
The contributions of this paper are~\textemdash
\begin{enumerate}
    \item shows that induced orbit counting algorithms can be extended to non-induced orbits with low overhead,
    \item introduces efficient, parallelisable strategies to accelerate computation,
    \item demonstrates strong scalability and runtime performance across real-world and synthetic datasets.
\end{enumerate}

The rest of this paper is organised as follows. Section~\ref{niorca:sec:related} reviews related work on graphlet and orbit counting algorithms. Section~\ref{niorca:sec:preliminaries} defines graphlets and orbits. Section~\ref{niorca:sec:orca_overview} details the original ORCA algorithm, followed by Section~\ref{niorca:sec:orca_to_niorca}, which presents the theoretical foundations and modifications required for NI-ORCA. Section~\ref{niorca:sec:parallelisation_strategies} outlines the parallelisation techniques developed for each algorithm stage. Section~\ref{niorca:sec:experimental_setup} describes the experimental setup, while Section~\ref{niorca:sec:experimental_evaluation} presents an evaluation across eight real-world and multiple synthetic datasets. Finally, Section~\ref{niorca:sec:conclusion} summarises the key findings and outlines directions for future work.

\section{Related Work}
\label{niorca:sec:related}
The two major types of graphlet analysis~\textemdash~induced and non-induced~\textemdash~differ significantly in resilience to noisy or incomplete data. It has been argued that non-induced counts are sufficient for many applications, especially when role-based analysis (via orbits) is the goal~\cite{ortmann2017efficient}. Algorithms like ESCAPE~\cite{pinar2017escape} count non-induced graphlets frequencies to discover graphlets efficiently. However, ESCAPE lacks per-vertex or per-edge orbit counting. Nevertheless, algorithms for graphlet counting have been extended for orbit counting. For example, RAGE~\cite{marcus2012rage} (Rapid Graphlet Enumerator) performs orbit counting by identifying the specific structural roles (automorphism orbits) that each vertex plays within small graphlets (up to size four) as they are enumerated in the network.

ORCA~\cite{hovcevar2014combinatorial} introduced an efficient system of combinatorial equations for computing induced graphlet orbits up to size five. This method was extended in EVOKE~\cite{pashanasangi2020efficiently}, which builds on ESCAPE, using a graph-cutting strategy to generalise orbit counting and improve performance in counting orbits of 5-vertex graphlets. While these methods are powerful, they are both designed for induced subgraphs and are sequential, making them ill-suited for real-world applications with missing or noisy edges. Moreover, EVOKE uses the count of orbits of 4-vertex graphlets in a preprocessing stage to efficiently count 5-vertex graphlets' orbits.

JESSE~\cite{melckenbeeck2016algorithm} enables orbit counting for graphlets of arbitrary size. The algorithm uses a hierarchical construction of graphlets using a directed tree structure. It employs automatically generated linear equations that relate the orbit counts of larger graphlets to those of smaller ones. This allows orbit degrees of graphlets of order $n$ to be computed by enumerating only graphlets of order $n-1$, reducing computational complexity. Additionally, the method uses symmetry-breaking constraints derived from stabiliser chains of automorphism groups to eliminate redundant isomorphic subgraphs, ensuring correctness and efficiency.

CORA~\cite{prvzulj2007biological} uses automorphism orbits to capture and compare the local topology of biological networks. Instead of relying on global metrics, the approach defines Graphlet Degree Distributions (GDDs) corresponding to the distinct vertex orbits found across all 2–5 node connected induced graphlets. Each GDD quantifies how frequently nodes in a network participate in specific structural roles defined by these orbits. By comparing the normalised GDDs of two networks, the method computes a similarity score~\textemdash~called the GDD-agreement~\textemdash~that reflects how similar the networks are in their local orbit-based topology. CORA is implemented for network similarity determination in the GraphCrunch~\cite{milenkovic2008graphcrunch,kuchaiev2011graphcrunch} tool.

While the BEACON benchmark~\cite{najafi2025beacon} and GraphCrunch emphasise the need for scalable orbit-based analysis, both focus exclusively on induced graphlets. BEACON categorised and evaluated both algorithmic (including exact methods ESCAPE and EVOKE, and the approximate method MOTIVO) and Graph Neural Network (GNN)-based methods (\cite{maron2019provably,qian2022ordered,schwabe2024cardinality}) by accuracy and performance. It confirms that no method simultaneously provides exact, scalable, and parallel support for non-induced orbit counting. It also observes that machine learning-based methods offer scalability at the cost of accuracy and reproducibility. 

DISC~\cite{zhang2020distributed} is a distributed orbit counting algorithm implemented using Spark. It uses homomorphism-based matching and relational algebra (e.g., joins, group-by, aggregation). However, its reliance on relaxed injectivity results in overcounting. It needs to convert overcounted (non-injective) homomorphic counts into exact counts of injective graphlets. This is corrected through Möbius inversion (a combinatorial technique to reverse a summation over subsets or partitions) and recursive filtering, making it inefficient for larger
or more complex graphlets.

However, a well-structured sequential algorithm can outperform a distributed algorithm, as shown by SCOPE efficiency over DISC in real-world datasets~\cite{li2024fast}. SCOPE is a tree decomposition-based approach to decompose a complex pattern graph into smaller components and count local subgraph matches more efficiently. It uses tISO (tree-based isomorphism), which balances homomorphism and subgraph isomorphism by enforcing isomorphic matching only within parts of the pattern. It employs symmetry-breaking rules to minimise redundant computations and a multi-join algorithm to process the tree decomposition efficiently.

\section{Preliminaries}
\label{niorca:sec:preliminaries}
\begin{figure}[ht]
\centering
\includegraphics[width=0.7\textwidth]{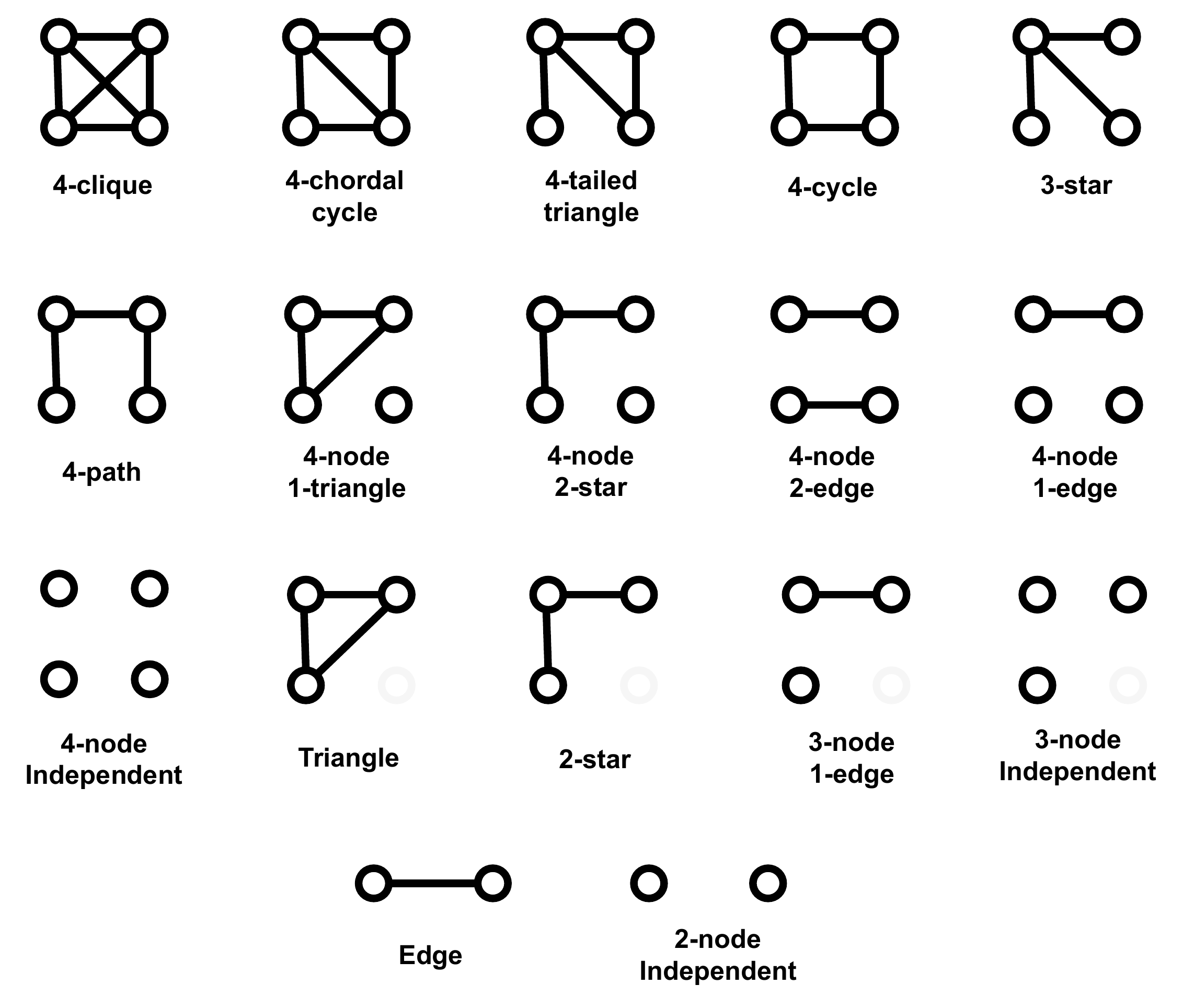}
\caption{Seventeen graphlets formed using up to four vertices - The nine connected graphlets are considered for further orbit definitions.}
\label{fig:niorca:graphlets}
\end{figure}

\begin{definition}[Graphlet]
Let $G\:=\:(V,\;E)$ be a simple graph with vertex set $V$ and edge set $E$. A graphlet $G'$ is an isomorphism class of subgraphs of $G$. Specifically, two occurrences of a graphlet $G'$ in $G$, say $ G'_1$ and $ G'_2$, are isomorphic, i.e., there exists a bijective mapping between the vertex sets $ G'_1$ and $ G'_2$ preserving adjacency. Similarly, two different graphlet classes of $G$, say $G'$ and $G''$, are non-isomorphic.
\end{definition}

\begin{definition}[Induced Graphlet]
Given graph $G$, an induced graphlet $G'\,\,=(V',\,E')$ is a connected subgraph of $G$ formed by a subset of nodes $V'\:{\subseteq}\:V$ along with all edges in $E$ that exist between nodes in $G$. Formally \textemdash
\begin{equation}
E' = \{ (u, v) \in E \mid \{u, v\} \in V' \}.
\end{equation}
\end{definition}

\begin{definition}[Non-Induced Graphlet]
Given graph $G$, a non-induced graphlet $G'\,\,=(V',\,E')$ is a connected subgraph of $G$ that consists of a subset of nodes $V'\:{\subseteq}\:V$ and a subset of edges $E'\:{\subseteq}\:E$ that connect nodes in $V'$. In this case, \( E' \) may include any subset of the edges between nodes in \( V' \), without requiring all such edges present in \( G \). Formally \textemdash

\begin{equation}
E' \subseteq \{ (u, v) \in E \mid u, v \in V' \}.
\end{equation}
\end{definition}
~\\
Induced graphlets require all edges between a chosen set of vertices, reflecting the exact local structure, while non-induced graphlets allow selective edge inclusion, capturing more general connection patterns among vertices.

\begin{definition}[Automorphism]
Let $G'\:=\:(V',\;E')$ be a graphlet in $G\:=(V,\;E)$ where $V'\:{\subseteq}\:V$ is the set of vertices and $E'\:{\subseteq}\:E$ is the set of edges $G'$. An automorphism of $G'$ is an isomorphism (a bijective mapping) from $V'$ onto itself preserving adjacency. Formally, the set of all automorphisms of $G'$ is denoted \textemdash

\begin{align}
    \text{Aut}(G') = \{ f : V' \to V' \mid &\; f \text{ is bijective } \nonumber \\
    &\land\; (x, y) \in E' \Rightarrow (f(x), f(y)) \in E' \}.
\end{align}
\end{definition}

\begin{figure}[ht]
\centering
\includegraphics[width=0.5\textwidth]{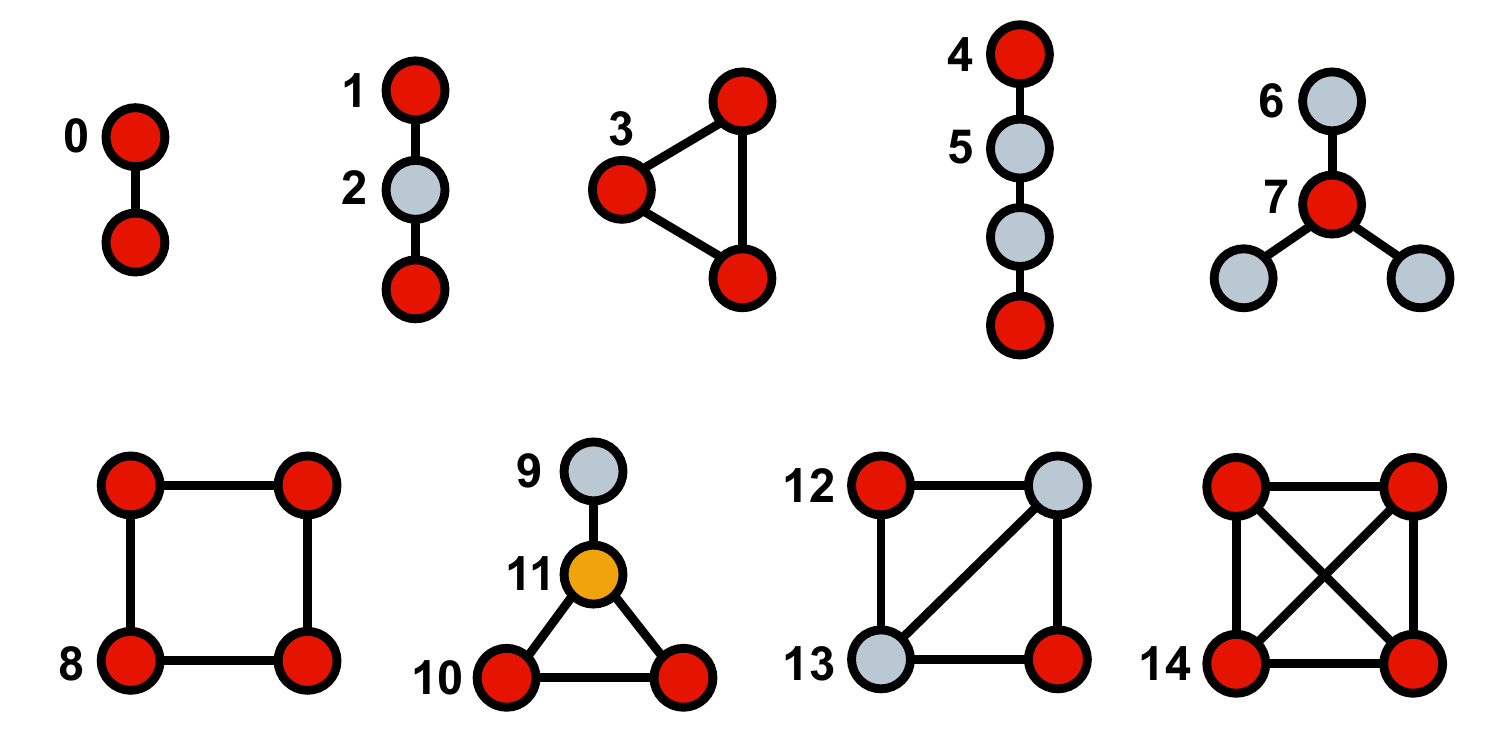}
\caption{Fifteen orbits formed using the nine connected graphlets - The colours in the vertices indicate their orbit roles within the graphlet, i.e., vertices with the same colour belong to the same orbit group.}
\label{fig:niorca:orbits}
\end{figure}

\begin{definition}[Orbit]
Given a node $v\:{\in}\:V'$, the orbit of $v$ under the action of $Aut(G')$, denoted $Orb(G',v)$, is the set of nodes in $G'$ that can be mapped to $v$ through any automorphism in $Aut(G')$. This set represents the collection of nodes in $G'$ that are symmetrically equivalent to $v$, meaning they occupy identical structural roles within the graphlet. Formally \textemdash

\begin{align}
    \text{Orb}(G', v) = \{ w \in V' \mid\; & \exists f \in \text{Aut}(G')\;\land\;f(v) = w \}
\end{align}
\end{definition}

The seventeen graphlets~\textemdash~including both connected and disconnected graphlets~\textemdash~that can be formed using up to four vertices are shown in Fig.~\ref{fig:niorca:graphlets}. For the purposes of orbit counting, this study focuses on the nine connected graphlets. These connected graphlets give rise to fifteen distinct orbits, each representing a unique automorphic role that a vertex can assume within a graphlet. These orbits are depicted in Fig.~\ref{fig:niorca:orbits}, where vertices sharing the same colour belong to the same orbit type, indicating their equivalence under the graphlet’s automorphism group. Disconnected graphlets are composed of smaller and previously accounted connected graphlets. As such, orbit counting over connected graphlets suffices to capture the complete set of structural roles of the vertices.

This study focuses on counting the orbits of non-induced graphlets with up to four vertices, i.e., up to $K_4$. For a graph, $G=(V,\:E)$, the complexity of counting orbits of 4-vertex graphlets is  $\mathcal{O}(|E|{\cdot}(\nicefrac{|E|}{|V|})+T_4)$, where $\mathcal{O}(T_4) = \mathcal{O}(|V|{\cdot}{(\nicefrac{|E|}{|V|})}^3)$ is the complexity of enumerating all $K_4$; whereas, for counting the orbits of 5-vertex graphlet, the complexity is $\mathcal{O}(|E|{\cdot}(\nicefrac{|E|}{|V|})^2+T_5)$, where $\mathcal{O}(T_5) = \mathcal{O}(|V|{\cdot}(\nicefrac{|E|}{|V|})^4)$ is the complexity of enumerating all $K_5$~\cite{hovcevar2014combinatorial}. Thus, counting the orbits of graphlets up to $K_5$ is an order of magnitude more computationally expensive than counting up to $K_4$~\textemdash~this added complexity might be prohibitive to many applications. The methods and observations in this study could easily be extended to count orbits up to $K_5$, which is beyond the scope of this paper. Additionally, recent algorithms for counting orbits up to $K_5$, like EVOKE~\cite{pashanasangi2020efficiently}, count orbits up to $K_4$ as a preprocessing step and could benefit from this study.

\section{Overview of the ORCA algorithm}
\label{niorca:sec:orca_overview}

This study builds on ORCA due to its foundational role in efficient orbit counting via a combinatorial, equation-based framework. ORCA computes induced graphlet orbits without complete subgraph enumeration. Its use of local structural variables and symbolic equation systems provides both interpretability and amenability to further optimisation. However, to the best of our knowledge, all existing methods, including ORCA, suffer from two critical gaps: (i) A lack of support for exact non-induced orbit counting across all graphlets, and (ii) the absence of scalable, parallel implementations capable of exploiting modern multi-core architectures.

The ORCA algorithm computes node orbit counts in graphlets up to size four (up to $K4$) through a three-stage process that avoids full enumeration and instead uses combinatorial relationships among subgraphs. In the first stage, the algorithm efficiently counts triangles for every edge in the graph. This provides common neighbours and triangle participation statistics, which are later reused in orbit calculations. In the second stage, the algorithm enumerates one orbit directly~\textemdash~the complete 4-node clique ($K4$, i.e., orbit 14)~\textemdash~using an optimised enumeration method based on intersecting neighbourhoods. In the third stage, ORCA constructs a linear equation system that relates the frequencies of the 11 orbits in all 4-vertex graphlets. These equations are based on how smaller 3-node subgraphs (like paths or triangles) involving a given vertex can be extended to 4-vertex graphlets by adding a fourth vertex, depending on edge configurations. This staged approach drastically reduces the computation time, making ORCA up to an order of magnitude faster than full enumeration methods. \textbf{We demonstrate that counting the orbits of non-induced graphlets can be structured as a similar computation to those of induced graphlets, with the difference being in~the~linear~equation~system}.

\begin{figure}[htbp!]
    \begin{subfigure}{.2\linewidth}
        \centering
        \includegraphics[width=0.6\linewidth]{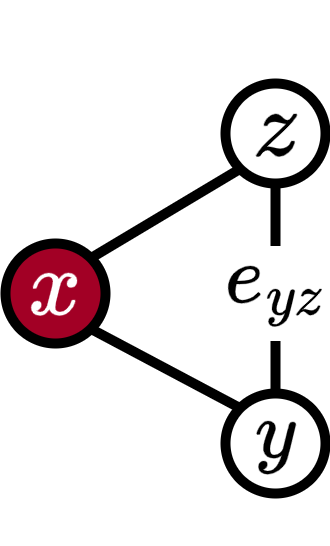}
        \caption{Triangle 1}
        \label{fig:orca_s3_fig_1}
    \end{subfigure}\hfill 
    \begin{subfigure}{.2\linewidth}
        \centering
        \includegraphics[width=0.6\linewidth]{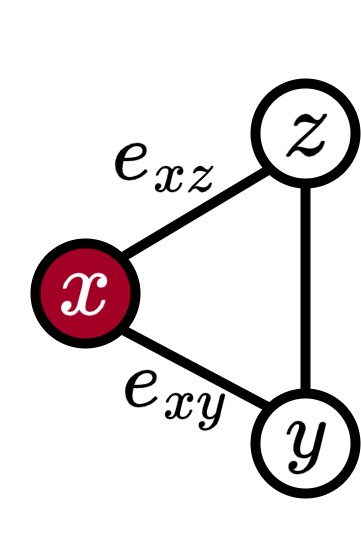}
        \caption{Triangle 2}
        \label{fig:orca_s3_fig_2}
    \end{subfigure}\hfill
    \begin{subfigure}{.2\linewidth}
        \centering
        \includegraphics[width=0.6\linewidth]{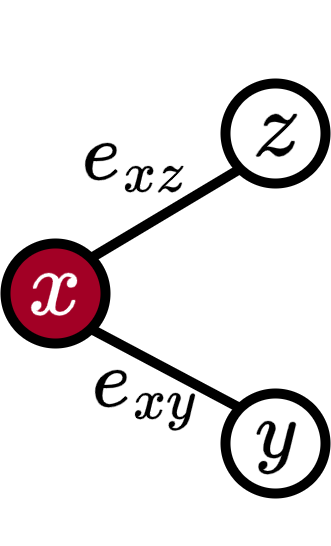}
        \caption{Two Star}
        \label{fig:orca_s3_fig_3}
    \end{subfigure}\hfill 
    \begin{subfigure}{.2\linewidth}
        \centering
        \includegraphics[width=0.6\linewidth]{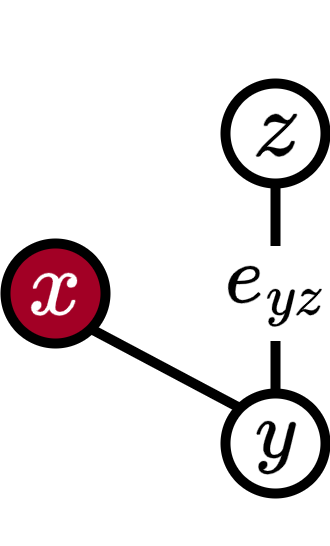}
        \caption{Path}
        \label{fig:orca_s3_fig_4}
    \end{subfigure}
    \caption{Structures around $x\:{\in}\:V$ to compute the $f*$ internal variables in Stage III}
    \label{fig:orca_s3_figs}
\end{figure}

As outlined in Algorithm~\ref{alg:orca}, the orbit counting process consists of three main stages, with the final stage further divided into two sub-stages. \textbf{Stage~I} focuses on counting \textit{edge triangles}, i.e., all three-vertex subgraphs where two vertices in an edge share a common neighbour. \textbf{Stage~II}, described in Algorithm~\ref{alg:countK4}, identifies and counts 4-cliques ($K_4$ subgraphs). This is done by examining each vertex and its neighbours: for a given pair of neighbouring vertices, the algorithm checks whether they share any common neighbours. If any pair among these common neighbours is also connected, then the four vertices—the original two plus the connected pair from their shared neighbours—form a 4-clique. In such a case, the count of $K_4$ is incremented for each of the four involved vertices.

\begin{algorithm}[htbp!]
\small
\caption{Stage II - Counting K4}
\label{alg:countK4}
\KwInput{Vertex $x\:{\in}\:V$, array \texttt{\textbf{K4}}}
\SetKwFunction{FCount}{CountK4}
\SetKwProg{Fn}{Function}{:}{}
\Fn{\FCount{$x$, \texttt{\textbf{K4}}}}{
    \texttt{\textbf{neigh} = []}\;
    \ForAll{$y\;{\in}\;\texttt{neighbour}(x)\;{\land}\;(y\text{.id}\:{\geq}\:x\text{.id})$}{
        \ForAll{$z\;{\in}\;\texttt{neighbour}(y)\;{\land}\;(z\text{.id}\:{\geq}\:y\text{.id})$}{
            \If{$(x,\,z)\;{\in}\;E$}{
                \texttt{\textbf{neigh}}.append($z$)\;
            }
        }
        \For{$i \leftarrow 0$ \KwTo \texttt{len(\textbf{neigh})}} {
            $z=$\texttt{\textbf{neigh}[}$i$\texttt{]}\;
            \For{$j \leftarrow i$ \KwTo \texttt{len(\textbf{neigh})}} {
                $z'=$\texttt{\textbf{neigh}[}$j$\texttt{]}\;
                \If{$(z,\:z')\;{\in}\;E$} {
                    \texttt{\textbf{K4}[}$x$\texttt{\textbf{]}}++; \texttt{\textbf{K4}[}$y$\texttt{\textbf{]}}++\;
                    \texttt{\textbf{K4}[}$z$\texttt{\textbf{]}}++; \texttt{\textbf{K4}[}$z'$\texttt{\textbf{]}}++\;
                }
            }
        }
    }
}
\end{algorithm}

\begin{algorithm}[htbp!]
\small
\caption{Overview of ORCA}
\label{alg:orca}
\KwInput{Graph, $G\;=\;(V,\;E)$}
\KwOutput{Orbit Counts, $\{o_i\;|\;i\:{\in}\:{\mathbb{Z}}\;{\land}\;i\:{\in}\:[0, 14]\}$}
\SetKwFunction{FORCA}{ORCA}
\SetKwProg{Fn}{Function}{:}{}
\Fn{\FORCA{}}{
    \Comment{Stage I - Count Per-Edge Triangles}
    \texttt{\textbf{tri} = []}\;
    \For{$i$ from $0$ to $|E|$} {
        $(x, y)\;{\gets}\;i^{\text{th}}$ edge of $E$\;
        \texttt{\textbf{tri}}$[i]$${\gets}\;\;|$\,\texttt{common\_neighbours}$(x,\:y)|$\;
    }
    \Comment{Stage II - Count $K_4$}
    \texttt{\textbf{K4} = []}\;
    \ForAll{$x\;{\in}\;V$} {
        \texttt{countK4}($x$, \texttt{\textbf{K4}})\;
    }
    \Comment{Stage III - Count orbits}
    \ForAll{$x\;{\in}\;V$} {
        \texttt{\textbf{common} = []}\;
        $f*\;=\;$\texttt{count\_f\_vars()} \tcp*{See Table~\ref{niorca:tab:internal_variables} for details}
        \texttt{solve\_equations()} \tcp*{See Section~\ref{niorca:sec:orca_to_niorca} for details}
    }
}
\end{algorithm}

\textbf{Stage~III} calculates detailed \textit{orbit counts}, which represent the roles a vertex plays within different graphlet structures. This stage consists of two sub-stages: In the \textit{first sub-stage}, the algorithm calculates nine internal variables, referred to as $f*$ variables. For a given vertex $x \in V$, the algorithm explores its local neighbourhood (including triangle configurations) to identify specific substructures, as illustrated in Fig.~\ref{fig:orca_s3_figs}. Among these, Fig.~\ref{fig:orca_s3_fig_1} shows the only ordered substructure, where the index of vertex $y$ is greater than that of $z$. While probing this substructure, if $(x, z) \notin E$, the ordering is not considered, and \texttt{\textbf{common[}}$z$\texttt{\textbf{]}} is used to record the number of times vertex $z$ appears in this configuration via different $y$ vertices for a fixed $x$ and $z$. In other words, \texttt{\textbf{common[}}$z$\texttt{\textbf{]}} records how many times $x$ and $z$ are connected by an induced 3-path. The full definitions of the~nine~internal~$f*$~variables~are provided in Table~\ref{niorca:tab:internal_variables}.

\begingroup
\setlength{\parskip}{1pt}
\renewcommand{\arraystretch}{1.2}
\linespread{1.0}\selectfont
\newcolumntype{L}[1]{>{\raggedright\arraybackslash}p{#1}}
\begin{longtable}{|L{0.15\textwidth}|L{0.15\textwidth}|L{0.58\textwidth}|}
\caption{Description of internal $f*$ variables} \label{niorca:tab:internal_variables} \\

\hline
\shortstack{\textbf{Internal}\\\textbf{Variable}} & 
\shortstack{\textbf{Reference}\\\textbf{Image}} & 
\shortstack{\textbf{Description}} \\
\hline
\endfirsthead

\hline
\shortstack{\textbf{Internal}\\\textbf{Variable}} & 
\shortstack{\textbf{Reference}\\\textbf{Image}} & 
\shortstack{\textbf{Description}} \\
\hline
\endhead

\hline
\endfoot

\hline
\endlastfoot

$f_{12}^{14}$ & Fig.~\ref{fig:orca_s3_fig_1} & Count of triangles of $e_{yz}$, except ${\Delta}xyz$ \\ \hline
$f_{10}^{13}$ & Fig.~\ref{fig:orca_s3_fig_1} & Count of neighbours of $y$ and $z$ that do not form a triangle with $e_{yz}$ \\ \hline
$f_{13}^{14}$ & Fig.~\ref{fig:orca_s3_fig_2} & Count of triangles formed by $e_{xy}$ and $e_{xz}$, excluding ${\Delta}xyz$ \\ \hline
$f_{11}^{13}$ & Fig.~\ref{fig:orca_s3_fig_2} & Count of neighbours of $x$ that do not form triangles with $y$ or $z$ \\ \hline
$f_{5}^{8}$ & Fig.~\ref{fig:orca_s3_fig_3} & Count of all the neighbours of $y$ and $z$ (excluding $x$) that do not form a triangle with $x$ \\ \hline
$f_{6}^{9}$ & Fig.~\ref{fig:orca_s3_fig_4} & All the neighbours of $y$ (except $x$ and $z$) that do not form a triangle with $e_{yz}$ \\ \hline
$f_{9}^{12}$ & Fig.~\ref{fig:orca_s3_fig_4} & Count of triangles formed by $e_{yz}$ \\ \hline
$f_{4}^{8}$ & Fig.~\ref{fig:orca_s3_fig_4} & Count of all neighbours of $z$ (excluding $y$ and those that form triangles with $y$) \\ \hline
$f_{8}^{12}$ & Fig.~\ref{fig:orca_s3_fig_4} & Number of times $z$ was in this configuration except through $y$ (\texttt{\textbf{common[}}$z$\texttt{\textbf{]}}-1) \\ \hline
\end{longtable}
\endgroup

In the \textit{second sub-stage}, these $f*$ variables are used in a system of fifteen equations~\textemdash~one per orbit~\textemdash~to calculate the orbit counts for each vertex. These equations are discussed in detail in Section~\ref{niorca:sec:orca_to_niorca}.

\section{ORCA to NI-ORCA}
\label{niorca:sec:orca_to_niorca}

Let ${\psi}_k(x)$ denote the count of any \textit{non-induced} orbit $o_k$ incident on a given vertex $x$. Some orbit counts can be computed directly from earlier stages of the algorithm: ${\psi}_0(x)$ corresponds to the degree of vertex $x$ and can be trivially obtained as ${\psi}_0(x)\:=\:$\texttt{\textbf{deg[}}$x$\texttt{\textbf{]}}. Similarly, ${\psi}_{14}(x)$, representing the 4-clique ($K_4$) orbit, is already computed in Stage II and retrieved as ${\psi}_{14}(x)\:=\:$\texttt{\textbf{K4[}}$x$\texttt{\textbf{]}}. The orbits for three-vertex graphlets~\textemdash~${\psi}_1(x)$ (side vertex of a 3-path), ${\psi}_2(x)$ (middle vertex of a 3-path), and ${\psi}_3(x)$ (triangle vertex)~\textemdash~are computed in the first substage of Stage III while probing substructures to count the $f^*$ internal variables. Finally, ${\psi}_7(x)$, which corresponds to the middle vertex of a 3-star, can be trivially~computed~as~a~binomial combination of its neighbours.

The remaining orbit counts for vertex $x$ are derived using a system of equations that involve the internal $f^*$ variables. We discuss four of these equations in the remainder of this section. The rest of the equations~are~provided~in~the~Appendix.

\begin{equation}
\label{eq:o13}
{\psi}_{13}(x)=\frac{f_{13}^{14}}{2}
\end{equation}

\begin{equation}
\label{eq:o12}
{\psi}_{12}(x)=f_{12}^{14}
\end{equation}

\begin{equation}
\label{eq:o11}
{\psi}_{11}(x)=\frac{f_{11}^{13} + f_{13}^{14}}{2}
\end{equation}

\begin{equation}
\label{eq:o10}
{\psi}_{10}(x)=f_{10}^{13}+2{\times}f_{12}^{14}
\end{equation}

Many of these equations include divisions or multiplications by factors of 2 to correctly account for symmetry or ordering in the substructures (e.g., swapping $y$ and $z$ in asymmetric configurations). For example, \ref{eq:o13} (counting ${\psi}_{13}(x)$ using Fig.~\ref{fig:orca_s3_fig_2}) is divided by 2 because it uses the variable $f^{14}_{13}$, which is based on an \textit{unordered} substructure. Swapping vertices $y$ and $z$ yields the same configuration, so the count must be halved to eliminate double counting. In contrast, ~\ref{eq:o12} (counting ${\psi}_{12}(x)$ using Fig.~\ref{fig:orca_s3_fig_1}) uses $f^{14}_{12}$, which is derived from an \textit{ordered} substructure. Since each configuration is unique, no division is necessary.

\ref{eq:o13} and \ref{eq:o12} are trivially true and can be interpreted straightforwardly. We discuss the correctness and interpretability of two of the non-trivial equations~\textemdash~\ref{eq:o11} involving division to eliminate double counting, and~\ref{eq:o10} that requires multiplication by two to introduce double counting. The remaining equations (given in the Appendix) are not elaborated on due to space limitations. However, they all follow similar reasoning and can be interpreted using the same symmetry principles and role assignment in the graphlet substructures.

For~\ref{eq:o11} (counting ${\psi}_{11}(x)$ using Fig.~\ref{fig:orca_s3_fig_2}), the term $f^{14}_{13}$ captures cases where $y$ is an $o_9$ vertex and triangles on $e_{xz}$ identify the two $o_{10}$ vertices (one of which is $z$). The configuration is symmetric, so it’s counted again when $y$ and $z$ are swapped. Similarly, $f^{13}_{11}$ captures cases where $y$ and $z$ are both $o_{10}$ vertices and $x$ is the $o_9$ vertex in the triangle $\Delta xyz$, again counted twice due to symmetry. Both terms are therefore divided by 2 to remove this redundancy.

Similarly, for \ref{eq:o10} (counting ${\psi}_{10}(x)$ using Fig.~\ref{fig:orca_s3_fig_1}), $f^{13}_{10}$ includes two ordered (and therefore, not redundant) configurations: one where $x$ and $y$ are $o_{10}$ vertices and $z$ is $o_{11}$, and one where $y$ and $z$ are swapped. $f^{14}_{12}$ is multiplied by 2 to account for both triangle configurations around edge $e_{yz}$. In both cases, the third triangle vertex $z'$ corresponds to $o_9$. Specifically, in one case $y$~is~the~other~$o_{10}$~vertex~and~$z$~is~$o_{11}$; in the other, $y$ and $z$ are swapped.

A key difference between \textbf{NI-ORCA} and the original ORCA algorithm is removing the need for subtractive corrections when computing orbit counts. In ORCA, since it counts \textit{induced} graphlets, the orbit count of larger subgraphs must be explicitly subtracted from the orbit counts of smaller overlapping subgraphs to avoid overcounting. Let ${\psi}'_k(x)$ denote the count of an \textit{induced} orbit $o_k$ incident on vertex $x$. For example, when counting ${\psi}'_{13}(x)$ the induced count of $o_{14}$ (the 4-clique) must be subtracted from the count of $o_{13}$ using the equation:

\begin{equation}
{\psi}'_{13}(x)=\frac{f_{13}^{14}}{2}-3\:{\times}\:{\psi}_{14}(x)
\end{equation}

Similarly, for the orbits $o_{10}$ and $o_{11}$ (which belong to the \textit{4-tailed triangle} graphlet), the induced ORCA method computes their counts using:

\begin{equation}
\label{orca:o11}
{\psi}'_{11}(x) = \frac{f_{11}^{13} - f_{13}^{14}}{2} + 3 \times f_{14}
\end{equation}

\begin{equation}
\label{orca:o10}
{\psi}'_{10}(x) = f_{10}^{13} - f_{13}^{14} + 6 \times f_{14}
\end{equation}

In contrast, \textbf{NI-ORCA}, which counts \textit{non-induced} graphlets, directly avoids such corrective subtractions. The modified equations in NI-ORCA for computing ${\psi}_{11}(x)$ and ${\psi}_{10}(x)$ (given in Equations~\ref{eq:o11} and~\ref{eq:o10}) do not require subtracting the influence of larger graphlets like $o_{14}$. Conceptually, the difference is analogous to the \textit{inclusion-exclusion principle}, where overlapping contributions from larger structures must be subtracted out. NI-ORCA inherently avoids these overlaps, allowing orbit counts to be computed more directly and efficiently.

\section{NI-ORCA: Parallelisation Strategies}
\label{niorca:sec:parallelisation_strategies}

As seen in Algorithm~\ref{alg:orca}, Stage I iterates the graph's edge list and counts the number of triangles in which an edge participates. The count is obtained by finding the cardinality of the set intersection of the neighbours of the two vertices of the edge. In each iteration, the count is stored in an array (i.e., \textbf{\texttt{tri}}) indexed by the edge's identifier. No contention exists among the different iterations in accessing the edge list. Therefore, the first stage can be trivially parallelised. 

However, unlike Stage I, parallelising the remaining two stages is non-trivial. In Stage II, described in Algorithm~\ref{alg:countK4}, array \texttt{\textbf{K4}} stores the count of $K_4$ for all vertices in the graph. Each iteration of this stage finds the count of $K_4$ for some vertex $x$. First, for a given $x$, each vertex $y$ such that edge $(x,\,y)\,{\in}\,E$ is evaluated to find all neighbours that form a triangle with $x$ and $y$. All such neighbours are stored in \texttt{\textbf{neigh}} array. Then, for each pair of vertices, $\{z, z'\}\;{\in}\;$\texttt{\textbf{neigh}}, if $(z,\,z')\,{\in}\,E$, it implies that $\{x,\:y,\:z,\:z'\}$ forms a four clique, i.e., $K_4$. Therefore, the counts in \texttt{\textbf{K4}} corresponding to the four vertices in the clique are incremented.

\begin{table}[t]
\setlength{\tabcolsep}{10pt} 
\renewcommand{\arraystretch}{1.2} 
\centering
\begin{tabularx}{\textwidth}{|l|X|}
\hline
\textbf{Mode} & \textbf{Description} \\ \hline
\texttt{\textbf{ARR}} & Each vertex has its private \texttt{\textbf{neigh}} array, and \texttt{\textbf{K4}} is atomically incremented in real-time whenever a $K_4$ is encountered. \\ \hline
\texttt{\textbf{UOM}} & Each thread has its private \texttt{\textbf{neigh}} and a local \texttt{\textbf{K4}} implemented using an Unordered Hash-Map (UOM). The global \texttt{\textbf{K4}} is incremented lazily at the end of the thread after all the local $K_4$ are counted. \\ \hline
\texttt{\textbf{FHM}} & Same as \texttt{\textbf{UOM}}, but \texttt{\textbf{neigh}} implemented using a Flat Hash-Map (FHM). \\ \hline
\end{tabularx}
\caption{Configuration modes for implementation of Stage II}
\label{tab:s2_mode}
\end{table}

Two data structures could potentially face contention in Stage II during parallelisation~\textemdash~\texttt{\textbf{K4}} and \texttt{\textbf{neigh}}. First, the \texttt{\textbf{K4}} data structure is accessed at four different indices corresponding to the four vertices for the 4-clique. This contention is removed by~\textemdash~(i) making all access to the \texttt{\textbf{K4}} \textbf{array} atomic, or (ii) buffering the \texttt{\textbf{K4}} locally in each thread and \textit{lazily} updating the global \texttt{\textbf{K4}} after the thread finishes execution. When parallelised using threads, each thread will not access the indices corresponding to all the vertices in the graph. Therefore, using an array as a buffer would be memory-inefficient. Therefore, the local buffer is implemented using an \textbf{Unordered Hash-Map} (UOM) or~a~\textbf{Flat~Hash-Map}~(FHM).

UOM is implemented by a typical chaining-based hash table, providing constant-time average-case insertions and lookups. However, it incurs overhead from dynamic memory allocation and pointer dereferencing, which can be a performance bottleneck when orbit counts are updated at scale. FHM, on the other hand, offers a more cache-friendly alternative by storing key-value pairs in contiguous memory blocks, significantly reducing access latency and improving spatial locality. This optimisation is particularly beneficial in high-thread-count environments, where contention is low but throughput demands are high.

Then, the contention for \texttt{\textbf{neigh}} array is removed by creating a local array per thread. This is initialised to be of size equal to the maximum degree in the graph. The array is reset at the start of each iteration (corresponding to a vertex) in a thread. Table~\ref{tab:s2_mode} describes the three modes of implementation of Stage II.

Finally, in Stage III of the algorithm, which involves probing local substructures around each vertex and solving the corresponding system of equations to compute orbit counts, may experience contention in the \texttt{\textbf{common}} data structure. This structure tracks the frequency of the number of times a vertex $z$ appears as a common neighbour between two vertices $x$ and $y$, used in computing the internal variable $f_{8}^{12}$. As each thread may simultaneously process multiple vertices and update their orbit counts, concurrent access to this~shared~data~structure~can~lead~to~race~conditions.

\bgroup
\begin{table}[htbp!]
\centering
\begin{tabular}{|l|l|l|}
\hline
\textbf{Mode} &  \textbf{Data Structure}  &  \textbf{Location} \\ \hline
\texttt{\textbf{ARR (V)}} &  Array & Per vertex \\ \hline
\texttt{\textbf{ARR (T)}} &  Array & Per thread \\ \hline
\texttt{\textbf{FHM (V)}} &  Flat Hash-Map & Per vertex \\ \hline
\texttt{\textbf{FHM (T)}} &  Flat Hash-Map & Per thread \\ \hline
\texttt{\textbf{UOM}} &  Unordered Map & Per vertex \\ \hline
\end{tabular}
\caption{Configuration modes for implementation of Stage III}
\label{tab:s3_mode}
\end{table}
\egroup

To mitigate this, the \texttt{\textbf{common}} structure can be replicated, similar to the strategy adopted in Stage II for \texttt{\textbf{neigh}} and \texttt{\textbf{K4}}. There are two primary approaches to replication: per-vertex and per-thread. In the per-vertex approach, each vertex in the graph maintains its own instance of the \texttt{\textbf{common}} data structure. While this eliminates contention entirely, it incurs a substantial memory overhead, especially in large graphs with millions of vertices. In contrast, the per-thread approach allocates a private instance of \texttt{\textbf{common}} for each thread. This approach offers a more memory-efficient alternative while still avoiding contention. However, it introduces a different form of overhead — each thread must perform additional housekeeping after processing each vertex to reset or clean up residual values in the \texttt{\textbf{common}} structure before processing the next vertex. This ensures stale data does not contaminate the results of subsequent computations.

Table~\ref{tab:s3_mode} summarises the five configuration modes for Stage III, categorised by both the type of data structure used (\texttt{\textbf{Array}}, \texttt{\textbf{Flat Hash-Map}}, \texttt{\textbf{Unordered Map}}) and its allocation scope (per vertex or per thread).

\section{Experimental Setup}
\label{niorca:sec:experimental_setup}

The experiments for this paper were performed in two systems. The first system, denoted \textbf{EPYC}, is powered by two AMD EPYC 7713 64-Core Processor. The system comprised 128 CPUs, distributed across two sockets with 64 cores each, running in a single-threaded configuration. It features eight NUMA nodes with a CPU clock speed of 1996.448 MHz. The cache hierarchy consisted of 32KB for both L1 data and instruction caches, 512KB for L2 cache, and 32MB for L3 cache.

The second system, denoted \textbf{XEON}, is powered by two Intel Xeon Gold 6438Y+ processors. The system consists of two sockets, each with 32 cores, supporting two threads per core with hyperthreading enabled, for a total of 128 usable CPUS. The CPU configuration is distributed across four NUMA nodes. The processors feature clock speeds ranging from 800 MHz to 4000 MHz. The cache architecture includes 3MB of L1 data cache, 2MB of L1 instruction cache, 128MB of L2 cache, and 120MB of L3 cache.

\bgroup
\def\arraystretch{1.2}
\begin{table}[htbp!]
    \centering
    \begin{tabular}{|l|c|c|c|c|}
    \hline
        \;\textbf{Dataset}\; & \textbf{$|V|$} & \textbf{$|E|$} & \textbf{$\mu$(d)} & \textbf{Density} \\ \hline
        \;\texttt{DBLP} & 317080 & 1049866 & 6.622 & 2.09E-05 \\ \hline
        \;\texttt{EU2005} & 862664 & 16138468 & 37.415 & 4.34E-05 \\ \hline
        \;\texttt{HPRD} & 9460 & 34998 & 7.399 & 7.82E-04 \\ \hline
        \;\texttt{HUMAN} & 4674 & 86282 & 36.92 & 7.90E-03 \\ \hline
        \;\texttt{PATENTS} & 3774768 & 16518947 & 8.752 & 2.32E-06 \\ \hline
        \;\texttt{WORDNET} & 76853 & 120399 & 3.133 & 4.08E-05 \\ \hline
        \;\texttt{YEAST} & 3112 & 12519 & 8.046 & 2.59E-03 \\ \hline
        \;\texttt{YOUTUBE} & 1134890 & 2987624 & 5.265 & 4.64E-06 \\ \hline
    \end{tabular}
\caption{Overview of the eight graphs}
\label{tab:niorca:datasets-overview}
\end{table}
\egroup

The datasets used in this study, shown in Table~\ref{tab:niorca:datasets-overview}, vary in both size and structure. Large datasets like \texttt{PATENTS}, \texttt{EU2005}, and \texttt{YOUTUBE} have millions of vertices and edges, while smaller datasets such as \texttt{YEAST}, \texttt{HUMAN}, and \texttt{HPRD} contain only a few thousand vertices and edges. The average degree ($\mu$(d)) also varies widely: \texttt{EU2005} and \texttt{HUMAN} exhibit high average degrees, showing dense connectivity, whereas \texttt{WORDNET} has a much lower average degree, reflecting a sparser network. These contrasts suggest significant differences in graph size, density and connectivity~patterns~across~the~datasets.

\begin{figure}[htbp!]
\centering
\includegraphics[width=0.7\linewidth]{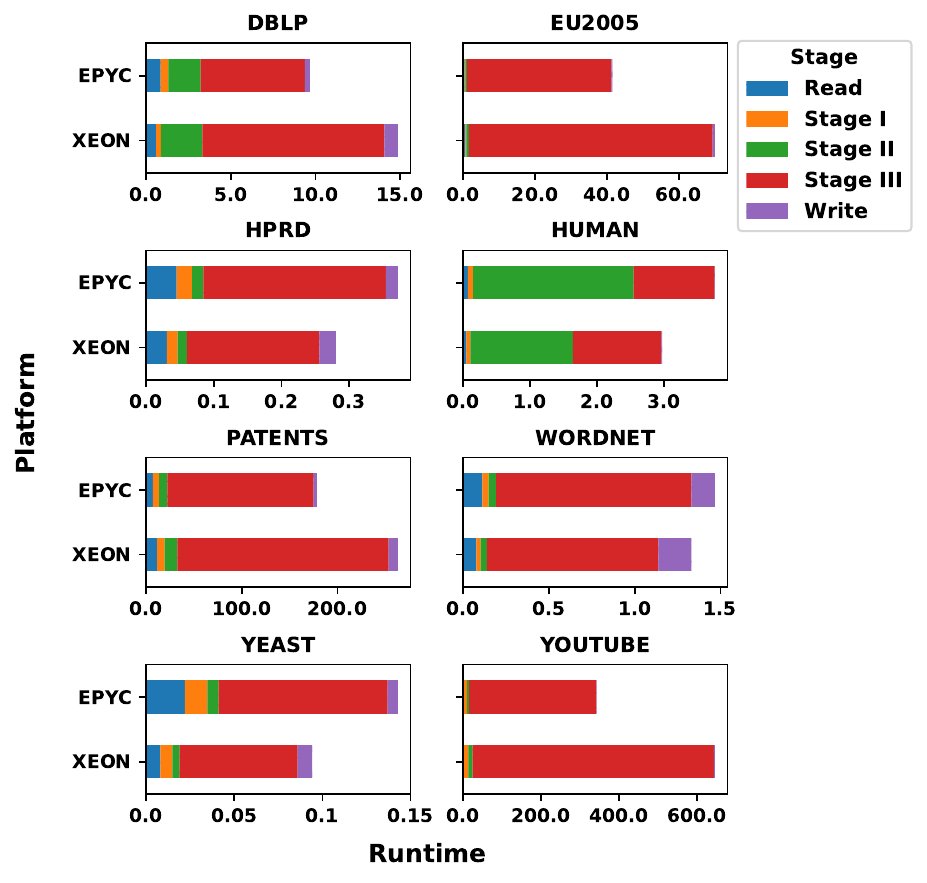}
\caption{Breakdown of the runtime of the three stages\\of the sequential ni-ORCA algorithm}
\label{fig:sequential_breakdown}
\end{figure}

While the study primarily focuses on real-world datasets, for analysis of runtime characteristic changes with the changing density of the graphs, an Erdős–Rényi random graph generator was used to generate multiple random graphs using NetworkX~\cite{hagberg2008exploring}. The random graphs can be categorised into two classes: (i) \texttt{SPARSE} (with $|V|\:{\approx}\:10^5$ and density between $10^{-8}$ to $10^{-2}$) and (ii) \texttt{DENSE} (with $|V|\:{\approx}\:10^3$ and density between $0.2$ to $0.8$). Three graphs were generated for each vertex count-density combination to normalise~the~runtime~characteristics~across multiple graphs.

\section{Experimental Evaluation}
\label{niorca:sec:experimental_evaluation}

While multiple studies have focused on efficient read and write I/O operations in graph analytics~\cite{esfahani2024selective,li2023graphar,li2022efficient}, addressing the preceding reading (the graph input file) and succeeding writing (the orbit count output file) stages of the algorithm is beyond the scope of this study. Instead, we focus only on the three core counting stages. The runtime breakdown of the sequential NI-ORCA algorithm is shown in Fig.~\ref{fig:sequential_breakdown}. Stage III (probing substructures and solving the system of equations) is the most expensive stage out of the algorithm's three core stages. Stage II's runtime follows that of Stage III (although Stage II is the most expensive stage in the \texttt{HUMAN} dataset). Finally, Stage I is the least expensive~of~the~three~core~stages.




\begin{figure}[ht]
\centering
\includegraphics[width=0.7\linewidth]{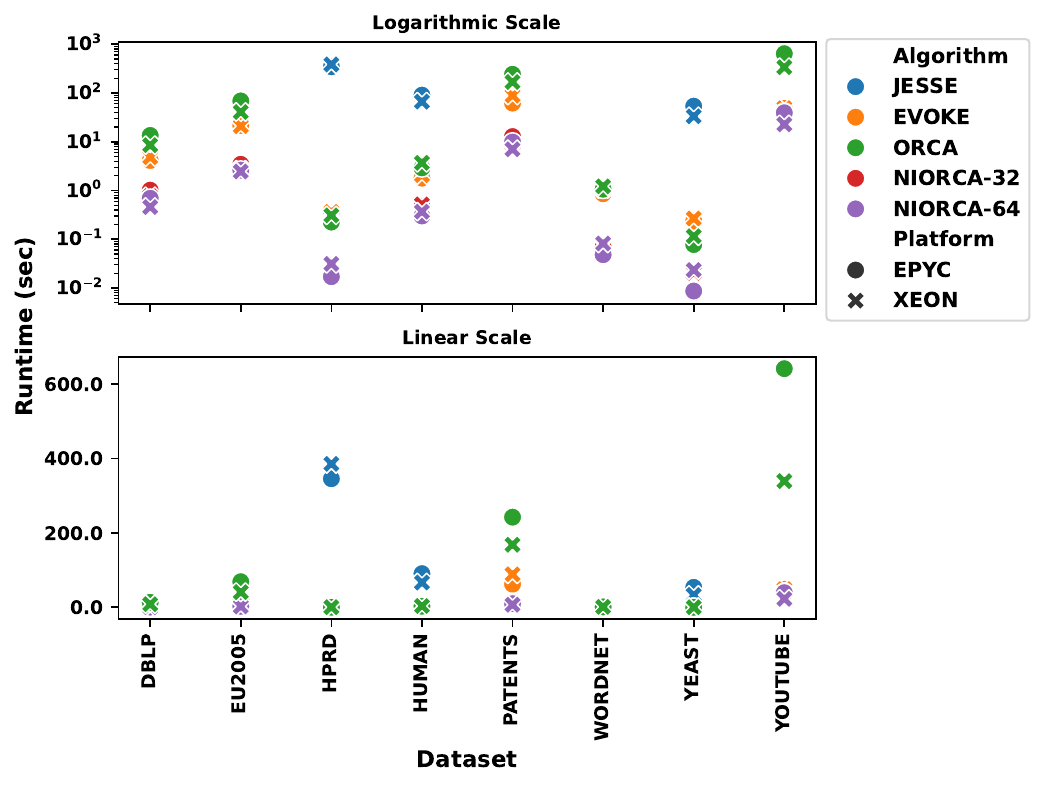}
\caption{Speedup using 32 and 64 threads of NI-ORCA compared to various baseline algorithms - JESSE completed execution successfully on only three datasets - CORA did not complete execution on any dataset within the threshold - EVOKE showed the best performance among all the sequential algorithms.}
\label{fig:mixed_speedup}
\end{figure}

Figure~\ref{fig:mixed_speedup} compares the runtime performance of NI-ORCA against state-of-the-art orbit counting algorithms~\textemdash~EVOKE\footnote{bitbucket.org/nojan-p/orbit-counting/}, ORCA\footnote{github.com/thocevar/orca}, JESSE\footnote{github.com/biointec/jesse/}, and CORA\footnote{github.com/benedekrozemberczki/OrbitalFeatures/}~\textemdash~across eight real-world datasets on the two platforms. The parallel implementations of NI-ORCA with 32 and 64 threads (\textbf{NIORCA-32} and \textbf{NIORCA-64}, respectively) consistently outperform all baseline algorithms. Among the sequential methods, EVOKE shows the best performance, but still lags significantly behind NI-ORCA in both thread configurations.

JESSE completes execution on only three datasets due to its reliance on an 
$n{\times}n$ Floyd–Warshall–based distance matrix computation, which results in excessive memory usage. Even on the EPYC platform with 512 GB of RAM, JESSE encounters out-of-memory (OOM) errors on larger graphs. CORA, which uses an enumeration-based orbit counting method, fails to complete execution within a one-hour threshold, even on small datasets like \texttt{YEAST} and \texttt{HPRD}.

While NIORCA-64 generally provides better speedup across larger and denser datasets~\textemdash~such as \texttt{PATENTS} and \texttt{EU2005}~\textemdash~the performance gains begin to plateau or even diminish slightly for smaller or sparser datasets like \texttt{YEAST}. This effect is attributed to increased thread management overhead and load imbalance at high thread counts. AMD EPYC consistently outperforms Intel XEON across most datasets and thread configurations among~the~two~hardware~platforms.

\begin{figure}[ht]
\centering
\includegraphics[width=0.7\linewidth]{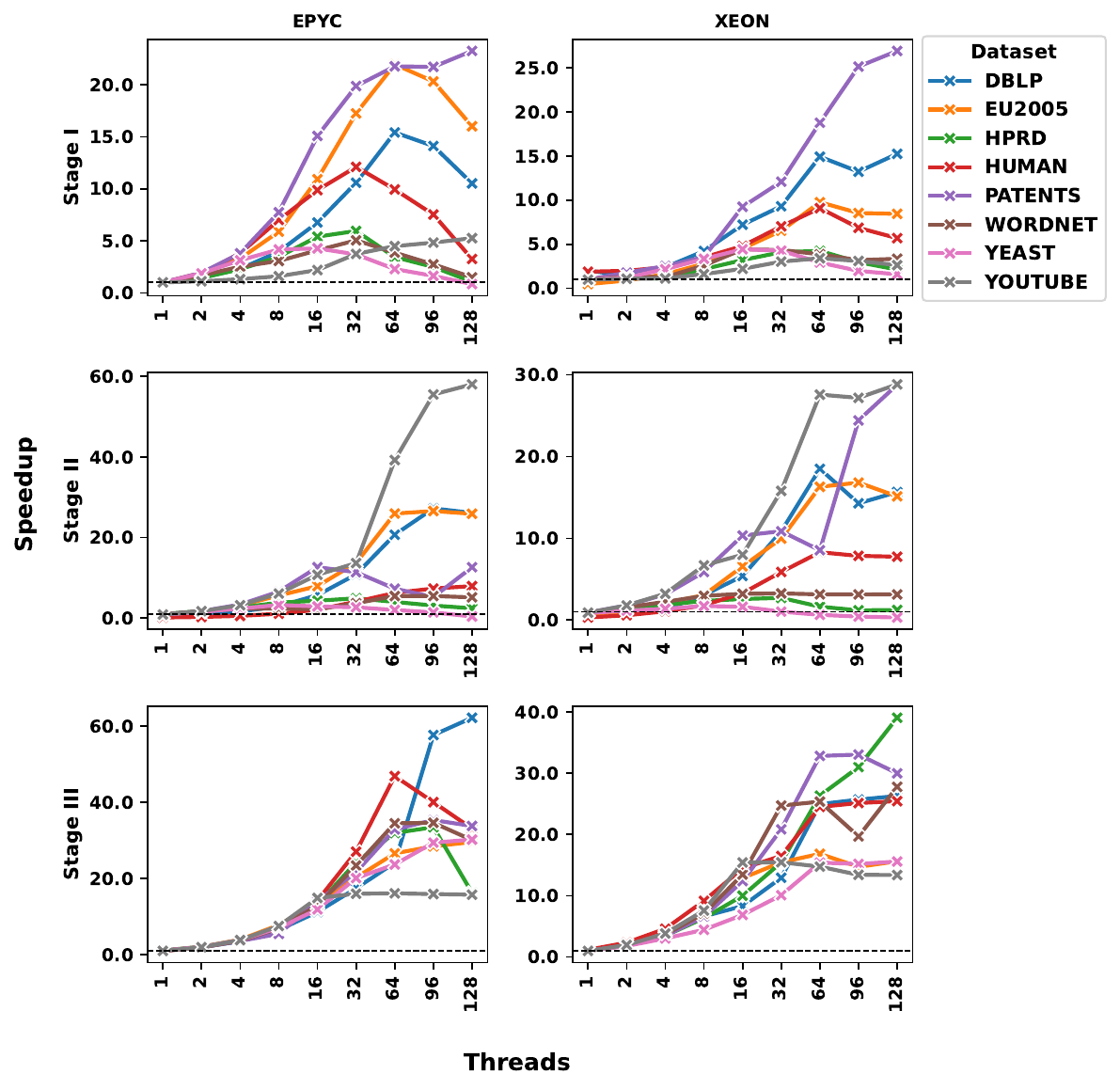}
\caption{Effect of parallelisation on Stage I - The dashed line corresponds to the runtime of the sequential algorithm.}
\label{fig:stage_1_speedup}
\end{figure}

\subsection{Stage-wise Runtime Profile of NI-ORCA}
The effect of parallelisation on the speedup of Stage I is shown in Fig.~\ref{fig:stage_1_speedup}. The speedup improves significantly as the thread count rises, demonstrating the efficiency of parallelisation for this stage. Each dataset shows different speedup ranges, but all follow a trend of improving speedup with additional threads. Similar trends can also be seen using the most optimised modes for Stage II and III. The figure demonstrates the scalability and efficiency of the algorithm’s parallel implementation across its three core stages. In the rest of this subsection, we discuss the influence of the choice of modes on the scalability of Stage II and III.

\begin{figure}[ht]
\centering
\includegraphics[width=0.7\linewidth]{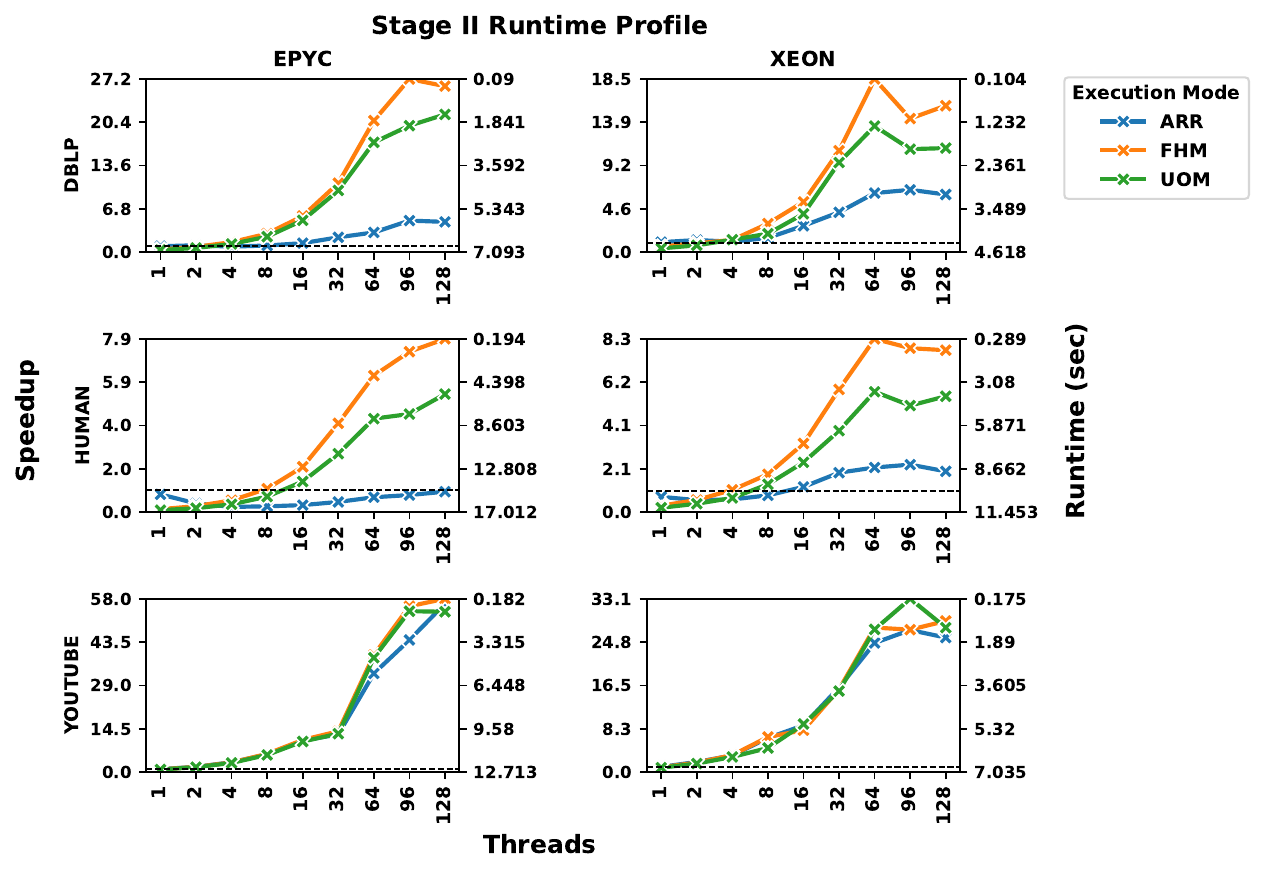}
\caption{Effect of parallelisation on Stage II~\textemdash~The dashed line (at $y=1$) corresponds to the runtime of the sequential algorithm.}
\label{fig:stage_2_speedup}
\end{figure}

Fig.~\ref{fig:stage_2_speedup} shows the influence of different parallelisation models on the speedup and runtime of Stage II of the algorithm. We focus on three datasets~\textemdash~\texttt{DBLP} and \texttt{HUMAN}, which,  in the sequential mode, had the most expensive relative runtimes in Stage II, and \texttt{YOUTUBE}, which had the highest overall runtime. The figure shows the influence of increasing thread count on runtime, comparing three parallelisation modes \textemdash~\textbf{\texttt{ARR}}, \textbf{\texttt{FHM}}, and \textbf{\texttt{UOM}} (described in Table~\ref{tab:s2_mode}). The speedup increases with increasing thread counts, demonstrating the efficiency gains from parallelisation. However, the benefits vary depending on the dataset, hardware platform, and specific parallelisation mode. The EPYC and XEON platforms show different runtime reductions due to their distinct architectures and multi-threading handling. Large datasets like \texttt{YOUTUBE} and \texttt{EU2005} (refer to Fig.~\ref{fig:stage_1_speedup}) exhibit more significant speedups with additional threads.

As the number of threads increases, the runtime generally decreases. The \textbf{\texttt{FHM}} (Flat Hash-Map) mode generally shows the best performance with its optimised structure, allowing for faster access and manipulation of data. While similar in structure, the \textbf{\texttt{UOM}} (Unordered Map) mode tends to lag slightly behind \textbf{\texttt{FHM}} due to its higher memory overhead, which is more pronounced in the \texttt{HUMAN} dataset. Additionally, the \textbf{\texttt{ARR}} (Array) mode demonstrates comparable performance to \textbf{\texttt{UOM}} in some cases (e.g., \texttt{PATENTS} and \texttt{YOUTUBE}), especially at lower thread counts. However, it significantly falls short in the \texttt{HUMAN} dataset. Thus, overall, \textbf{\texttt{FHM}} showed the most effectiveness across diverse datasets, especially at high thread counts. This is further corroborated by the other datasets not illustrated in the figure.

\begin{figure}[ht]
\centering
\includegraphics[width=0.7\linewidth]{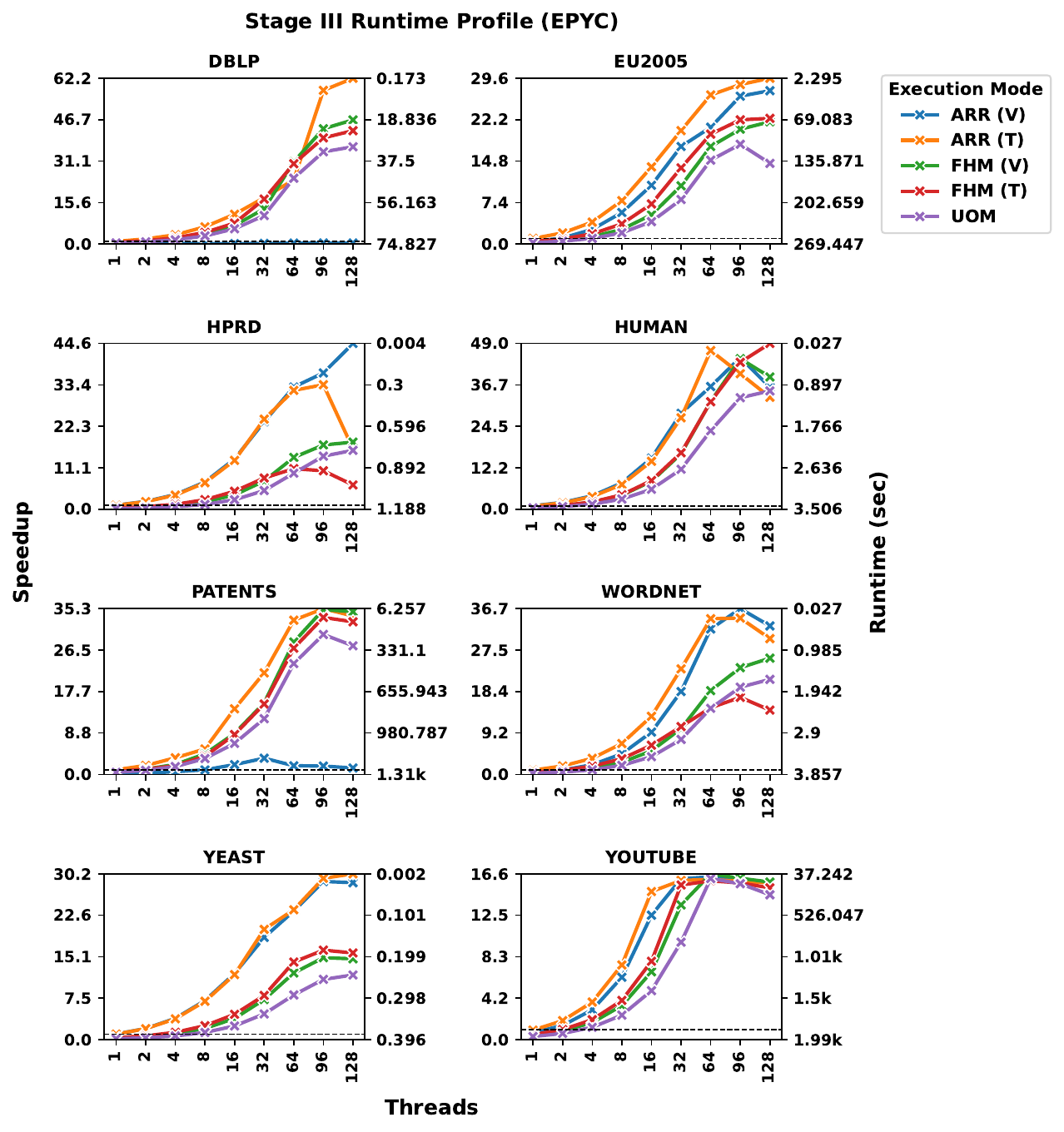}
\caption{Effect of parallelisation on Stage III on EPYC platform~\textemdash~The dashed line (at $y=1$) corresponds to the runtime of the sequential algorithm.}
\label{fig:stage_3_speedup}
\end{figure}

Fig.~\ref{fig:stage_3_speedup} shows the speedup comparison of Stage III using five modes \textemdash~\textbf{\texttt{ARR (T)}}, \textbf{\texttt{ARR (V)}}, \textbf{\texttt{FHM (T)}}, \textbf{\texttt{FHM (V)}}, and \textbf{\texttt{UOM}} (described in Table~\ref{tab:s3_mode}) on the EPYC platform. Among the different modes, \texttt{\textbf{ARR (T)}} consistently demonstrates the best performance, achieving the highest speedup across most of the datasets. While \texttt{\textbf{ARR (V)}} occasionally shows comparable performance, it shows the worst performance in datasets like \texttt{DBLP} and \texttt{PATENTS}. The \texttt{\textbf{UOM}} mode, on the other hand, consistently exhibits the worst performance overall, as it incurs higher memory overhead and slower access times. Among the flat hash-map approaches, \texttt{\textbf{FHM (T)}} generally performs better than \texttt{\textbf{FHM (V)}}, showing ever higher speedup that \texttt{\textbf{ARR (T)}} in \texttt{DBLP} at lower thread counts. Overall, \texttt{\textbf{ARR (T)}} is the most effective mode across datasets, especially with high thread counts.

\begin{figure}[ht]
\centering
\includegraphics[width=0.7\linewidth]{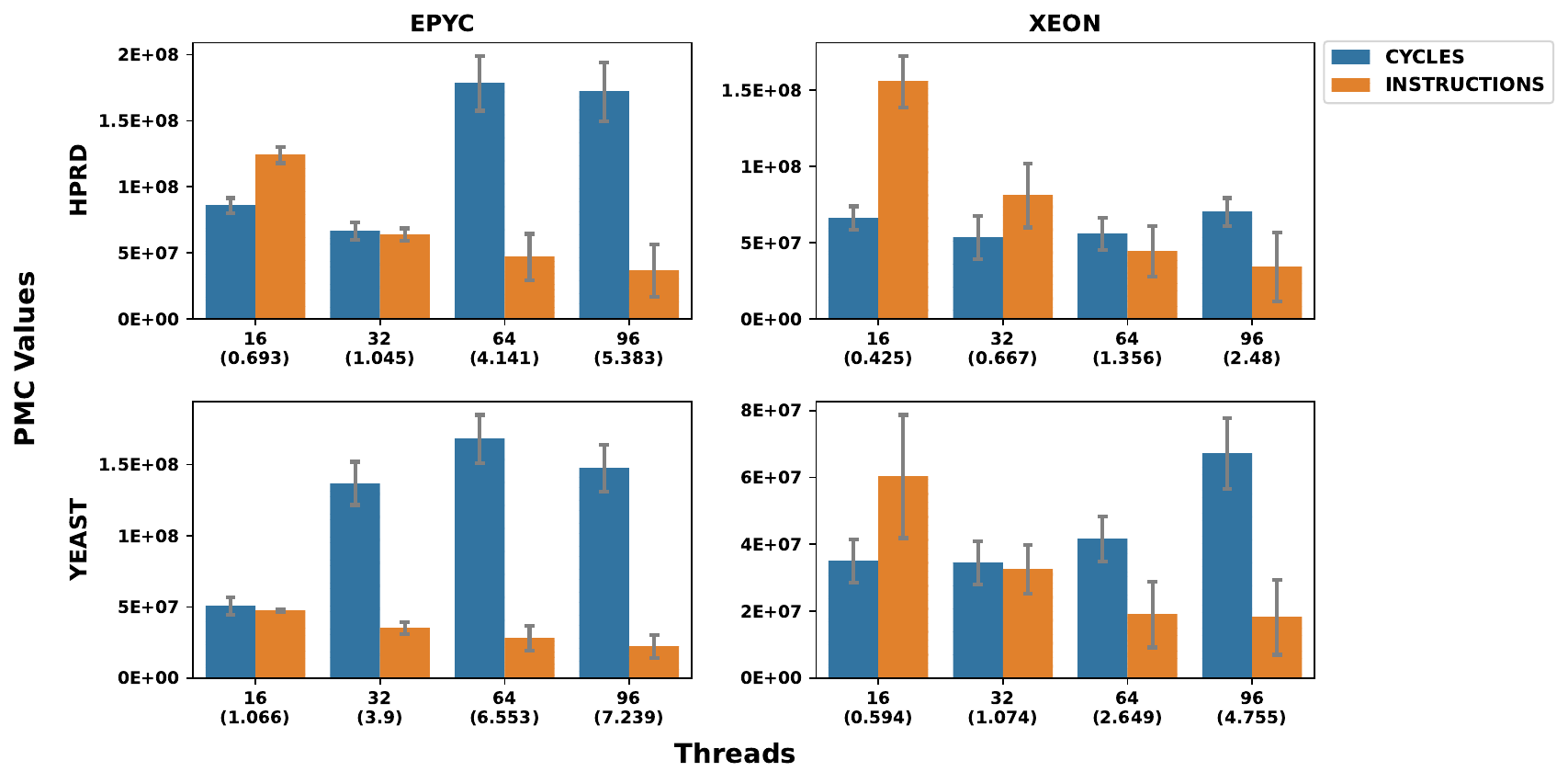}
\caption{Average per thread \texttt{Total Cycles} and \texttt{Total Instructions} performance monitoring counters profiled using PAPI - Figures in brackets indicate Cycles Per Instruction. Plots on the left are evaluated on the \texttt{EPYC} platform and those on the right on the \texttt{XEON} platform. Plots at the top correspond to the \texttt{HPRD} dataset, and those on the bottom correspond to the \texttt{YEAST} dataset.}
\label{fig:papi-statistics}
\end{figure}

The speedup achieved by NI-ORCA begins to plateau~\textemdash~and even deteriorates~\textemdash~at higher thread counts in smaller graphs like \texttt{HPRD} and \texttt{YEAST}, on both the \texttt{EPYC} and \texttt{XEON} platforms. The PAPI~\cite{danalis2022performance} performance counters, visualised in Fig.~\ref{fig:papi-statistics}, explain this trend. A consistently increasing discrepancy is seen between the number of cycles and the number of instructions executed per thread as thread counts increase from 16 to 96. For example, in \texttt{HPRD}, the Cycles per Instruction (CPI) on \texttt{EPYC} degrades from approximately 0.69 at 16 threads to 4.69 at 96 threads. Similar degradations can be seen in all other cases.

Simultaneously, the standard deviation in cycles and instruction counts increases, indicating worsening load imbalance across threads. These shifts suggest that although more threads are available, the computational workload becomes increasingly fragmented, leading to underutilisation of resources. This is manually verified by observing long-tailed behaviour through \texttt{htop}. This issue is more severe in \texttt{YEAST}, reflecting its smaller size and lower structural complexity.

The degraded scalability in small graphs arises because the available parallelism begins to exceed the computational workload, making tasks too fine-grained to offset the overheads of task scheduling, context switching, and memory synchronisation. Additionally, the rising per-thread disparity in performance metrics points to growing contention for shared caches and memory bandwidth, as well as NUMA-related inefficiencies in multi-socket systems.

\subsection{Scheduling Impact in Mixed Parallelisation}
The \textbf{\texttt{MIXED}} mode is a hybrid parallelisation strategy designed by combining \textbf{\texttt{FHM}} mode for Stage II with the \texttt{\textbf{ARR (T)}} mode for Stage III. The \textbf{\texttt{FHM}} mode minimises contention in memory-intensive operations during the counting of $K_4$. In contrast, the \textbf{\texttt{ARR (T)}} mode removes the memory management tasks required for per-vertex memory allocation. Fig.~\ref{fig:mixed_speedup} shows the efficient scalability of the \textbf{\texttt{MIXED}} mode across the eight datasets using various scheduling algorithms. The results demonstrate consistent speedups, particularly for large datasets like \texttt{PATENTS}, \texttt{EU2005}, \texttt{DBLP} and \texttt{YOUTUBE}, with the gains being more pronounced on the EPYC platform.

\begin{figure}[ht]
\centering
\includegraphics[width=0.9\linewidth]{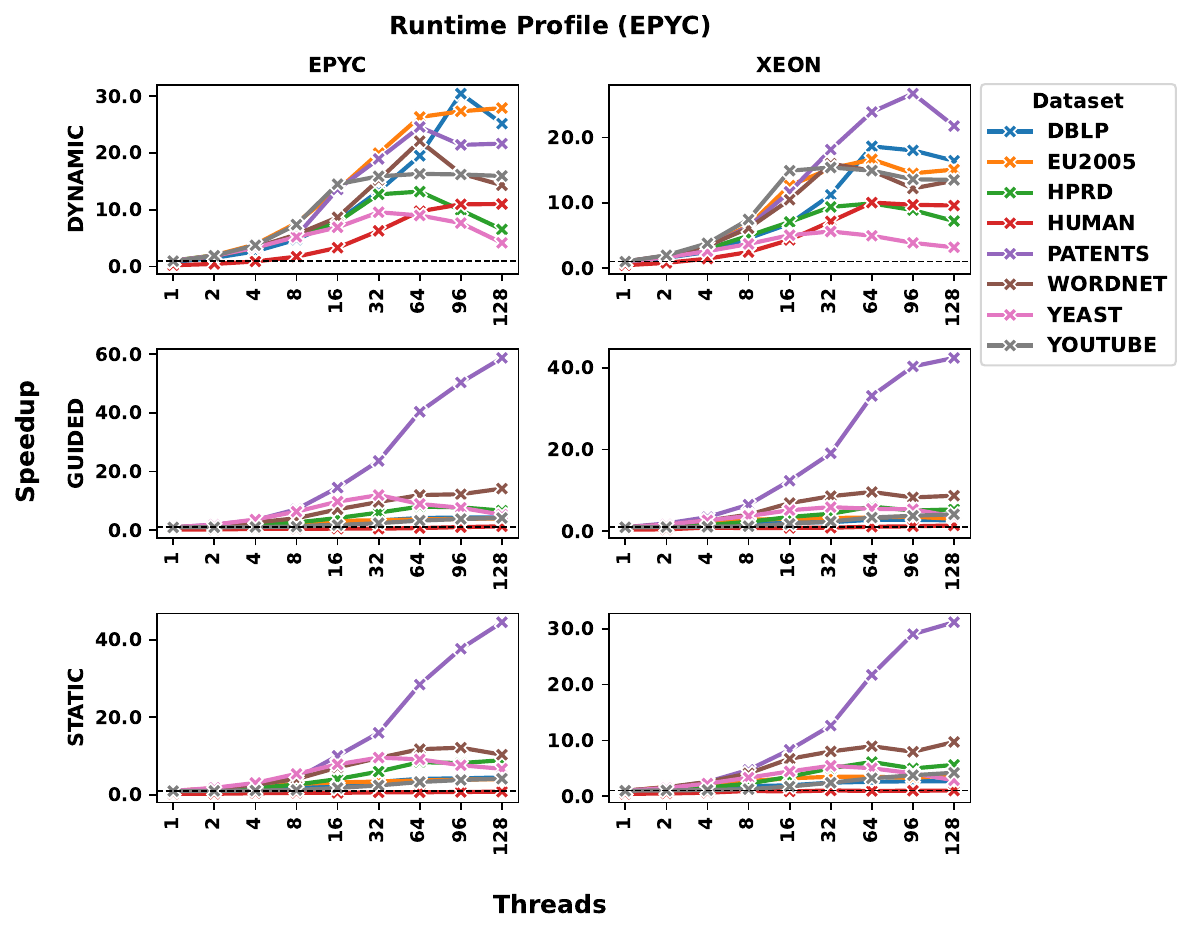}
\caption{Speedup using different scheduling algorithms using MIXED mode}
\label{fig:scheduling_impact}
\end{figure}

Across almost all the datasets, OpenMP's \textbf{dynamic} scheduling consistently outperforms \textbf{guided} and \textbf{static} scheduling methods in terms of speedup. Dynamic scheduling is beneficial in handling uneven workloads by dynamically distributing tasks among threads, resulting in better load balancing and reduced idle time. Although both guided and static scheduling performed outstanding scalability on the \texttt{PATENTS} dataset, it is an exception. In general, both tend to underperform compared to dynamic scheduling. In dynamic scheduling, this is because the chunk sizes in guided scheduling are large initially and only decrease over time, leading to less efficient use of threads for datasets with varying task complexities. Static scheduling shows the worst, as its predefined task distribution fails to account for differences in task durations, resulting in load imbalances hindering performance.

\begin{figure}[ht]
\centering
\includegraphics[width=0.8\linewidth]{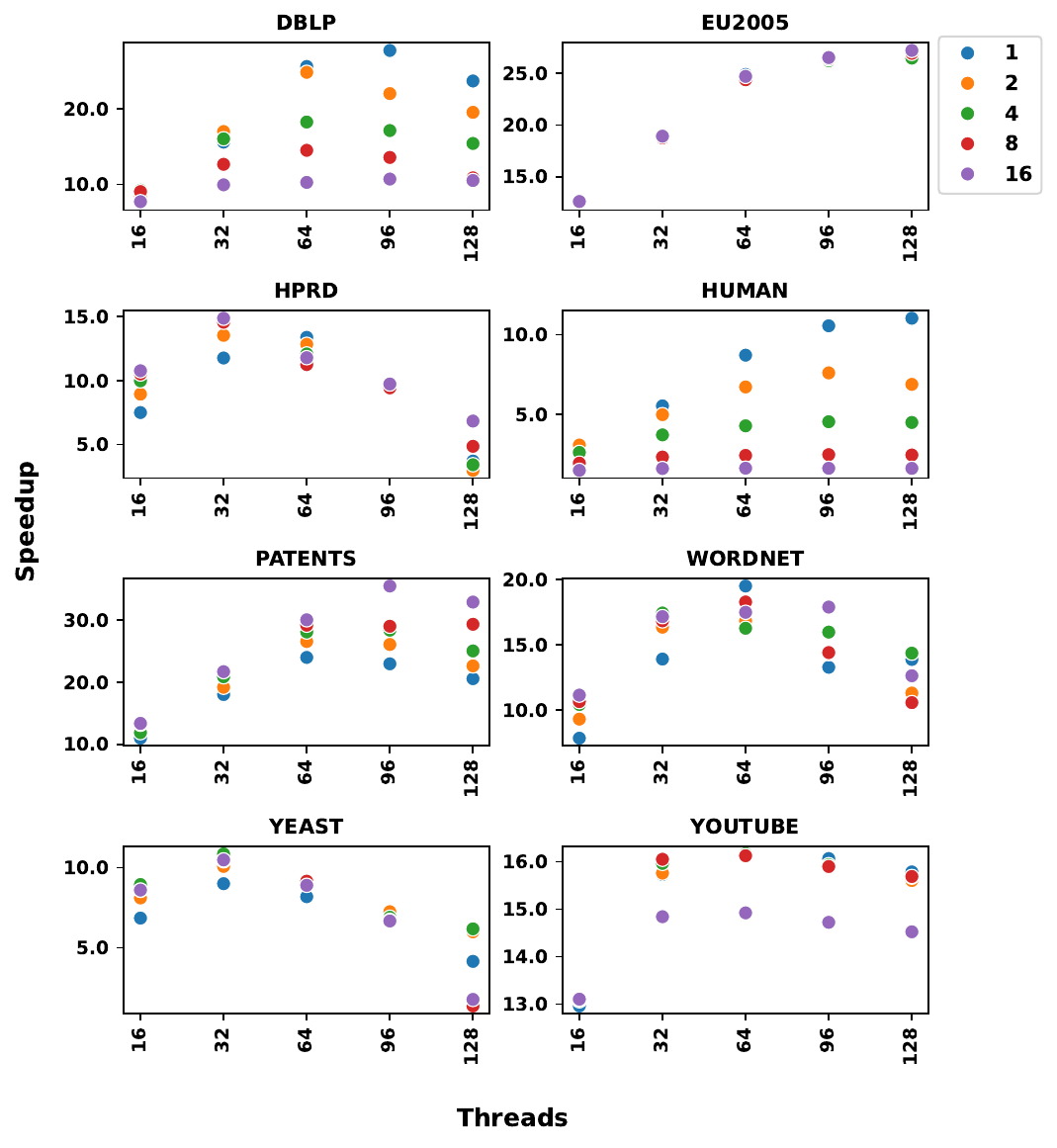}
\caption{Comparison of speedup for dynamic scheduling and various chunk sizes on the EPYC platform}
\label{fig:chunk_analysis}
\end{figure}

Fig.~\ref{fig:chunk_analysis} shows the performance of different chunk sizes in dynamic scheduling across the eight datasets in the EPYC platform. Larger chunk sizes yielded better performance for datasets like \texttt{PATENTS} and \texttt{WORDNET}. Here, the overhead of managing many small chunks outweighs the benefits of dynamic task allocation, resulting in diminishing returns on runtime reduction. However, smaller chunk sizes exhibited better performance in most cases. The performance improvement from smaller chunk sizes is more pronounced for datasets such as \texttt{DBLP} and \texttt{HUMAN}. Moreover, in \texttt{HUMAN}, the runtime improves as the thread count increases with a chunk size of one. However, in some datasets like \texttt{EU2005} and \texttt{YOUTUBE}, the performance improvement is independent of the chunk size, suggesting uniformity in per-vertex tasks resulting in uniform per-thread tasks.

\begin{figure}[ht]
\centering
\includegraphics[width=0.8\linewidth]{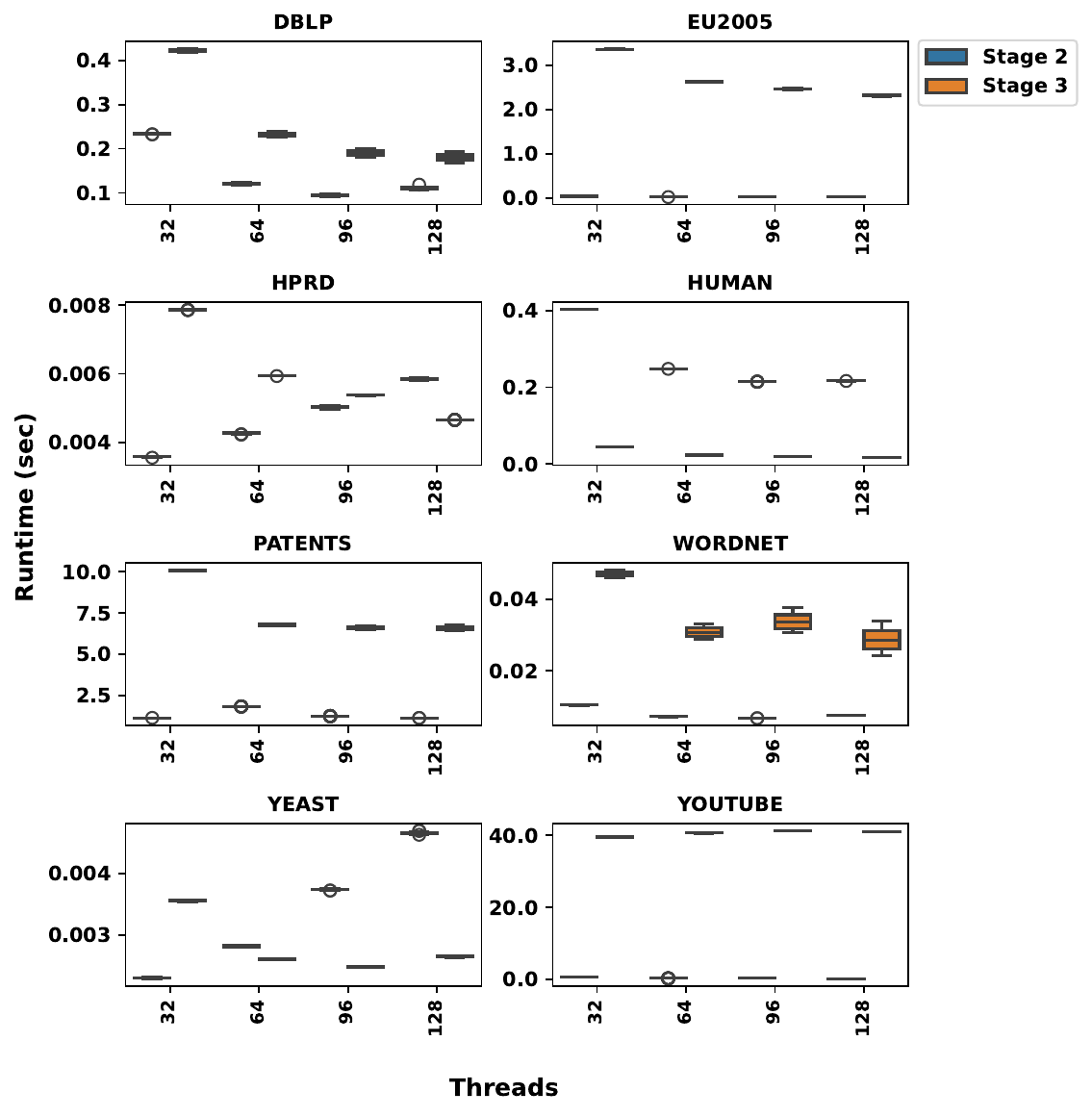}
\caption{Boxplots showing variation of runtime across all the threads Stage II and Stage III when running with multiple thread counts on the EPYC platform}
\label{fig:per-thread-analysis}
\end{figure}

Fig.~\ref{fig:per-thread-analysis} shows the variation in runtime across threads when dynamic scheduling is used. The low variations across all the datasets demonstrate the effectiveness of load balancing among the threads using dynamic scheduling. For both Stage II and Stage III, the near-uniform runtime per thread observed in all the datasets indicates that tasks are distributed efficiently, minimising idle time and ensuring optimal utilisation of computational resources. This highlights the adaptability of dynamic scheduling in managing diverse workloads by redistributing tasks dynamically as threads become available. While large datasets like \texttt{PATENTS}, \texttt{YOUTUBE} and \texttt{EU2005} with larger per-thread runtime do show low variation, even for smaller datasets like \texttt{HPRD} and \texttt{YEAST}, the consistent per-thread runtime demonstrates that the overhead of dynamic scheduling does not significantly impact efficiency. The increasing variance in the case of \texttt{WORDNET} dataset implies the uneven computational loads on different threads. This results in degraded performance at higher thread counts, as previously observed in Fig.~\ref{fig:scheduling_impact}.

\subsection{Scalability Across Graph Density and Size}
\begin{figure}[htbp!]
\centering
\includegraphics[width=0.7\linewidth]{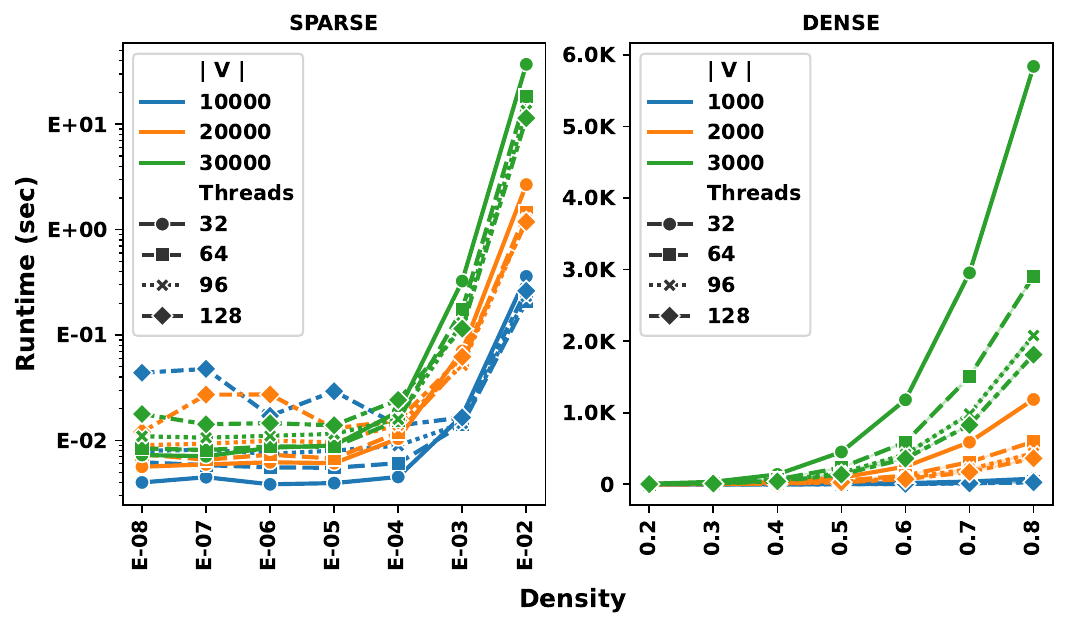}
\caption{Effect of parallelisation on random Erdős–Rényi graphs}
\label{fig:E3_density_sparse}
\end{figure}

The runtime of the parallelised NI-ORCA algorithm on randomly generated graphs with varying densities and vertex counts are shown in Fig.~\ref{fig:E3_density_sparse}. For sparse graphs, the runtime does not exhibit a strictly increasing trend as density increases from extremely low values ($10^{-8}$) to moderate values ($10^{-4}$). The trend fluctuations indicate that thread management overshadows the algorithm's runtime in sufficiently small (sparse) graphs. However, as the density increases from moderate values ($10^{-4}$) to moderately high values ($10^{-2}$), the fluctuations vanish and the runtime significantly increases with increasing density. As expected, larger graphs with 30,000 vertices take longer to process than smaller graphs with 10,000 or 20,000 vertices due to the higher computational demand. The runtime improves with an increasing number of threads, but the benefits diminish at very low densities, where the parallelisation overhead outweighs performance gains.

For dense graphs, runtime increases cubically with graph density (0.2 to 0.8) as additional edges increase computational complexity. The runtime consistently decreases as the number of threads increases, demonstrating the scalability of the proposed NI-ORCA algorithm. However, smaller graphs with 1,000 vertices exhibit less pronounced improvements due to the parallelism overhead.

The computational workload requires less processing in sparse graphs, where edges are significantly less. With fewer threads (e.g., 32), the overhead associated with thread management is minimal. On the other hand, at higher thread counts (e.g., 96 or 128), where synchronisation costs and resource contention dominate, it leads to diminishing performance. In contrast, dense graphs have a much substantial computational workload. With fewer threads, the algorithm underutilises the available computational resources. As the thread count increases, the workload is effectively parallelised to reduce runtime significantly. The transition in the scalability of the parallelisation occurs between moderate ($10^{-4}$) to moderately high densities ($10^{-2}$) in the sparse graphs. In~dense~graphs, the NI-ORCA~algorithm consistently~shows~effective scalability.

\section{Conclusion}
\label{niorca:sec:conclusion}

In this paper, we present \textbf{NI-ORCA}, a parallel algorithm for efficiently counting the orbits of non-induced graphlets up to $K_4$. The algorithm enhances the original ORCA algorithm by incorporating support for non-induced graphlets and applying parallelisation techniques to improve runtime performance.

Experimental evaluation across eight real-world and several synthetic datasets demonstrated that the algorithm scales well with increasing graph size and density. A \texttt{MIXED} parallelisation mode combining Flat Hash-Map (\texttt{FHM}) for $K_4$ counting and thread-local arrays (\texttt{ARR (T)}) for orbit computations provided the best runtime performance. Dynamic scheduling with small chunk sizes demonstrated efficient load balancing across threads, minimising idle time. The results also emphasise the scalability of NI-ORCA, particularly in dense graphs and large-scale datasets, where parallelisation provided consistent and substantial speedups.

The evaluation also demonstrated the impact of hardware architecture on performance. NI-ORCA showed significant speedup on multi-socket systems with optimised memory access, such as AMD EPYC and Intel XEON platforms, by effectively balancing workloads across NUMA nodes. However, it was observed that the performance plateaued at high thread counts in the sparse real-world and random graphs due to the increasing cost of thread synchronisation. The issue does not arise in dense random graphs, demonstrating~the~algorithm's~scalability~with~increasing~density.

NI-ORCA's scalable design and support for non-induced graphlets make it a valuable tool for analysing complex graphs in various domains. Future research could extend the algorithm to larger graphlets ($K_5$ and beyond), integrate it into distributed computing environments, and extend it to GPUs or other~accelerators~to~improve~performance further.

\newpage
\section*{Appendix (NI-ORCA)}
The equations for the rest of the orbits are as follows~\textemdash

\begin{equation}
\label{eq:o9}
{\psi}_9(x)=f_{12}^{14} + \frac{f_{9}^{12}}{2}
\end{equation}

\begin{equation}
\label{eq:o8}
{\psi}_8(x)=\frac{f_{13}^{14}+f_{8}^{12}}{2}
\end{equation}

\begin{equation}
\label{eq:o7}
{\psi}_7(x)=\binom{{\psi}_0(x)}{3}
\end{equation}

\begin{equation}
{\psi}_6(x)=\frac{f_{6}^{9}-f_{13}^{14}}{2}+f_{10}^{13}+2{\times}f_{12}^{14}
\end{equation}

\begin{equation}
{\psi}_5(x)=2{\times}f_{12}^{14}+f_{5}^{8}+f_{10}^{13}+f_{11}^{13}
\end{equation}

\begin{equation}
{\psi}_4(x)=2{\times}f_{12}^{14}+f_{4}^{8}+f_{10}^{13}+f_{9}^{12}
\end{equation}
\bibliographystyle{unsrt}  
\bibliography{references}
\end{document}